\documentclass{siamltex}
\usepackage{amsmath,amssymb,amsfonts,times,enumerate}
\usepackage{epsfig,hyperref}

\newtheorem{step}{Step}
\newtheorem{Prob}{Problem}

\DeclareMathOperator{\maj}{maj}
\newcommand{\mb}[1]{\ensuremath{\mathbf{#1}}}
\newcommand{\eat}[1]{}
\newcommand{\ann }[0]{\ensuremath{\wedge}}
\newcommand{\orr }[0]{\ensuremath{\vee}}
\newcommand{\easy}{tight}

\newcommand{\adj}{faithful}
\newcommand{\adv}{faithfully}

\newcommand{\FET}{Faithful Expressibility Theorem}

\newcommand{\cnf}{\mbox{\sc CNF}}
\newcommand{\stconn}{{\sc st-Conn}}
\newcommand{\conn}{{\sc Conn}}
\newcommand{\PTIME}{{\rm P}}
\newcommand{\NP}{{\rm NP}}
\newcommand{\coNP}{{\rm coNP}}
\newcommand{\PSPACE}{{\rm PSPACE}}

\newcommand{\sat}{{\sc Sat}}

\newcommand{\OR}{{\rm OR}}
\newcommand{\NAND}{{\rm NAND}}

\newcommand{\qaccept}{\ensuremath{q_{\rm{accept}}}}
\newcommand{\qreject}{\ensuremath{q_{\rm{reject}}}}

\newcommand{\comment}[1]{}

\title{The Connectivity of Boolean Satisfiability: \\
Computational and Structural Dichotomies}

\author{Parikshit Gopalan\thanks{University of Washington ({\tt
parik@cs.washington.edu}); work done in part while this author was
a summer intern at IBM Almaden.}
\and Phokion G.\ Kolaitis\thanks{IBM Almaden ({\tt
kolaitis@almaden.ibm.com}); on leave from UC Santa Cruz.}
\and Elitza  Maneva\thanks{IBM Almaden ({\tt
enmaneva@us.ibm.com})
; work done in part while this author was an intern at IBM Almaden.} 
\and ~~~~~~~~~~~~~~~~~~~~~~~~~~~~~~~~~~
Christos H. Papadimitriou\thanks{ UC Berkeley ({\tt
christos@cs.berkeley.edu}). Christos Papadimitriou's research was
supported by NSF grant CCF0635319, a gift from Yahoo! and a MICRO
grant, and a grant from the France-Berkeley Fund. }}

\begin{document}

\maketitle

\begin{abstract}
Boolean satisfiability problems are an important benchmark for
questions about complexity, algorithms, heuristics and threshold
phenomena. Recent work on heuristics, and the satisfiability threshold
has centered around the structure and connectivity of the solution
space.  Motivated by this work, we study structural and
connectivity-related properties of the space of solutions of Boolean
satisfiability problems and establish various dichotomies in
Schaefer's framework.

On the structural side, we obtain dichotomies for the kinds of
subgraphs of the hypercube that can be induced by the solutions of
Boolean formulas, as well as for the diameter of the connected
components of the solution space. On the computational side, we
establish dichotomy theorems for the complexity of the connectivity
and $st$-connectivity questions for the graph of solutions of Boolean
formulas.  Our results assert that the intractable side of the
computational dichotomies is \PSPACE-complete, while the tractable
side - which includes but is not limited to all problems with
polynomial time algorithms for satisfiability - is in \PTIME~for the
$st$-connectivity question, and in \coNP~for the connectivity
question. The diameter of components can be exponential for the
\PSPACE-complete cases, whereas in all other cases it is linear; thus,
small diameter and tractability of the connectivity problems are
remarkably aligned. The crux of our results is an expressibility
theorem showing that in the tractable cases, the subgraphs induced by
the solution space possess certain good structural properties, whereas
in the intractable cases, the subgraphs can be arbitrary.

\end{abstract}



\pagestyle{myheadings} \thispagestyle{plain} \markboth{P. GOPALAN,
P.G. KOLAITIS, E. MANEVA,  C.H. PAPADIMITRIOU}{THE CONNECTIVITY OF
BOOLEAN SATISFIABILITY}

\section{Introduction}

In 1978, T.J.\ Schaefer \cite{Sc78} introduced a rich framework for
expressing variants of Boolean satisfiability and proved a remarkable
{\em dichotomy theorem\/}: the satisfiability problem is in \PTIME~for
certain classes of Boolean formulas, while it is
\NP-complete for all other classes in the framework. In a single
stroke, this result pinpoints the computational complexity of all
well-known variants of \sat, such as $3$-\sat, {\sc Horn 3-\sat}, {\sc
Not-All-Equal $3$-\sat}, and {\sc $1$-in-$3$ Sat}. Schaefer's work
paved the way for a series of investigations establishing dichotomies
for several aspects of satisfiability, including optimization
\cite{Cr95,CKS01,KSTW01}, counting \cite{CH96}, inverse
satisfiability \cite{KS98}, minimal satisfiability \cite{KK03},
$3$-valued satisfiability \cite{Bu02} and propositional abduction \cite{abduction}.

Our aim in this paper is to carry out a comprehensive exploration of a
different aspect of Boolean satisfiability, namely, the {\emph
connectivity properties} of the space of solutions of Boolean
formulas. The solutions (satisfying assignments) of a given
$n$-variable Boolean formula $\varphi$ induce a subgraph $G(\varphi)$
of the $n$-dimensional hypercube. Thus, the following two decision
problems, called the \emph{connectivity problem} and the
\emph{$st$-connectivity problem}, arise naturally: (i) Given a Boolean
formula $\varphi$, is $G(\varphi)$ connected?  (ii) Given a Boolean
formula $\varphi$ and two solutions $\mb{s}$ and $\mb{t}$ of
$\varphi$, is there a path from $\mb{s}$ to $\mb{t}$ in $G(\varphi)$?

We believe that connectivity properties of Boolean satisfiability
merit study in their own right, as they shed light on the structure of
the solution space of Boolean formulas. Moreover, in recent years the
structure of the space of solutions for random instances has been the
main consideration at the basis of both algorithms for and
mathematical analysis of the satisfiability problem
\cite{ANP05,MZ02,MPZ02,MMW07}. It has been conjectured for
3-\sat~\cite{MPZ02} and proved for 8-\sat~\cite{MMZ05,AR06}, that the
solution space fractures as one approaches the {\em critical region}
from below. This apparently leads to performance deterioration of the
standard satisfiability algorithms, such as WalkSAT \cite{SKC93} and
DPLL \cite{ABM04}. It is also the main consideration behind the design
of the survey propagation algorithm, which has far superior
performance on random instances of satisfiability \cite{MPZ02}. This
body of work has served as a motivation to us for pursuing the
investigation reported here. While there has been an intensive study
of the structure of the solution space of Boolean satisfiability
problems for random instances, our work seems to be the first to
explore this issue from a worst-case viewpoint.

Our first main result is a dichotomy theorem for the $st$-connectivity
problem. This result reveals that the tractable side is much more
generous than the tractable side for satisfiability, while the
intractable side is \PSPACE-complete. Specifically, Schaefer showed
that the satisfiability problem is solvable in polynomial time
precisely for formulas built from Boolean relations all of which are
bijunctive, or all of which are Horn, or all of which are dual Horn,
or all of which are affine.  We identify new classes of Boolean
relations, called {\emph \easy}~relations, that properly contain the
classes of bijunctive, Horn, dual Horn, and affine relations. We show
that $st$-connectivity is solvable in linear time for formulas built
from tight relations, and \PSPACE-complete in all other cases. Our
second main result is a dichotomy theorem for the connectivity
problem: it is in \coNP~for formulas built from tight relations, and
\PSPACE-complete in all other cases.

In addition to these two complexity-theoretic dichotomies, we
establish a structural dichotomy theorem for the diameter of the
connected components of the solution space of Boolean formulas. This
result asserts that, in the \PSPACE-complete cases, the diameter of
the connected components can be exponential, but in all other cases it
is linear. Thus, small diameter and tractability of the
$st$-connectivity problem are remarkably aligned.

To establish these results,
the main 
challenge is to
show that for non-\easy~relations, both the connectivity problem
and the $st$-connectivity problem are \PSPACE-hard. In Schaefer's
Dichotomy Theorem, \NP-hardness of satisfiability was a
consequence of an \emph{expressibility} theorem, which asserted
that every Boolean relation can be obtained as a projection  over
a formula built from clauses in the ``hard" relations. Schaefer's
notion of expressibility is inadequate for our problem. Instead,
we introduce and work with a delicate and stricter notion of
expressibility, which we call \emph{\adj~expressibility}.
Intuitively, \adj~expressibility means that, in addition to
definability via a projection,  the space of witnesses of the
existential quantifiers in the projection has certain strong
connectivity properties that allow us to capture the graph
structure of the relation that is being defined. It should be
noted that Schaefer's Dichotomy Theorem can also be proved using a
Galois connection and Post's celebrated classification of the
lattice of Boolean clones (see \cite{BCRV04}).  This method,
however, does not appear to apply to connectivity, as the
boundaries discovered here cut across Boolean clones. Thus, the
use of \adj~expressibility or some other refined definability
technique seems unavoidable.

The first step towards proving \PSPACE-completeness is to show that
both connectivity and st-connectivity are hard for 3-\cnf~formulae;
this is proved by a reduction from a generic
\PSPACE~computation. Next, we identify the simplest relations that are
not tight: these are ternary relations whose graph is a path of length
$4$ between assignments at Hamming distance $2$. We show that these
paths can \adv~express all 3-\cnf~clauses. The crux of our hardness
result is an {\em expressibility} theorem
%phk showing
to the effect that one can \adv~express such a path from any set
of relations which is not tight. \eat{The property that the
distance between some pair of solutions expands plays a crucial
role in this expressibility theorem.   }

Finally, we show that all
\emph{\easy}~relations have ``good" structural properties.
Specifically, in a tight relation every component has a unique
minimum element, or every component has  a unique maximum element,
or the Hamming distance coincides with the shortest-path distance
in the relation.  These  properties are inherited by every formula
built from \easy~relations, and  yield both small diameter and
linear algorithms for $st$-connectivity.

%%%%%%%%%%%%%%%%%%%%%%%%%%%%%%%%%%%%%%%%%%%%%%%%%%%%%%%%%%%%%%%%%

Our original hope was that tractability results for connectivity could
conceivably inform heuristic algorithms for satisfiability and enhance
their effectiveness. In this context, our findings are {\em prima
facie\/} negative: we show that when satisfiability is intractable,
then connectivity is also intractable.  But our results do contain a
glimmer of hope: there are broad classes of intractable satisfiability
problems, those built from \easy~relations, with polynomial
$st$-connectivity and small diameter.
%phk It  is interesting to ask
It would be interesting to investigate if these properties make
random instances built from \easy~relations easier for
%heuristics
%like WalkSAT,
WalkSAT and similar heuristics, and if so, whether such heuristics
are amenable to rigorous analysis.

An extended abstract of this paper appears
in ICALP'06 \cite{icalp}.

\section{Basic Concepts and Statements of Results}
\label{sec:basic}

%phk added definition of CNF and k-CNF
A CNF formula is a Boolean formula of the form $C_1\land \dots
\land C_n$, where each $C_i$ is a clause, i.e.,  a disjunction of
literals.  If $k$ is a positive integer, then a $k$-CNF formula is
a CNF formula $C_1\land \dots \land C_n$ in which each clause
$C_i$ is a disjunction of at most $k$ literals.

 A {\em logical
relation\/} $R$ is a non-empty subset of $\{0,1\}^k$, for some
$k\geq 1$; $k$ is the {\em arity\/} of $R$. Let ${\cal S}$ be a
finite set of logical relations. A {\em \cnf$({\cal
S})$-formula\/} over a set of variables $V= \{x_1, \dots, x_n\}$
is a finite conjunction $C_1\land \dots \land C_n$ of clauses
built using relations from ${\cal S}$, variables from $V$, and the
constants $0$ and $1$; this means that each $C_i$ is an expression
of the form $R(\xi_1,\dots,\xi_k)$, where $R \in {\cal S}$ is a
relation of arity $k$, and each $\xi_j$ is a variable in $V$ or
one of the constants $0$, $1$.
%phk added definition of solution
A \emph{solution} of a \cnf$({\cal S})$-formula $\varphi$ is an
assignment $s=(a_1,\ldots,a_n)$ of Boolean values to the variables
that makes every clause of $\varphi$ true. A \cnf$({\cal
S})$-formula is \emph{satisfiable} if it has at least one
solution.

The \emph{satisfiability problem} \sat$({\cal S})$ associated with a
finite set ${\cal S}$ of logical relations asks: given a
\cnf$({\cal S})$-formula $\varphi$, is it satisfiable?
All well known restrictions of Boolean satisfiability, such as
{\sc $3$-\sat}, {\sc Not-All-Equal $3$-\sat}, and {\sc Positive $1$-in-$3$
\sat}, can be cast as \sat$({\cal S})$ problems, for a suitable choice of
${\cal S}$. For instance, let
$R_0=\{0,1\}^3\backslash\{000\}$,
$R_1=\{0,1\}^3\backslash\{100\}$,
$R_2=\{0,1\}^3\backslash\{110\}$,
$R_3=\{0,1\}^3\backslash\{111\}$. Then  {\sc $3$-\sat} is the problem 
\sat$(\{R_0,R_1, R_2, R_3\})$.
%phk this does not belong here, so I commented it out.
%For brevity, we will refer to the formulas corresponding to {\sc
%$k$-\sat} as $k$-CNF formulas.
%bringing back the pos.-1-in-3-sat example
Similarly, {\sc Positive 1-in-3\sat} is \sat$(\{R_{1/3}\})$, where
$R_{1/3}=\{100,010, 001\}$.

Schaefer \cite{Sc78} identified the complexity of \emph{every}
satisfiability problem \sat$({\cal S})$, where ${\cal S}$ ranges
over all finite sets of logical relations. To state Schaefer's
main result, we need to define some basic concepts.
\begin{definition}
\label{Schaefer} %{\rm  
Let $R$ be a logical relation.
\begin{enumerate}
\item $R$ is {\em bijunctive\/} if it is  the set of
solutions of a 2-\cnf~formula.
\item $R$ is {\em Horn\/} if it is the set of solutions
of a Horn formula, where a Horn formula is a \cnf~formula such
that each conjunct has at most one positive literal.
\item $R$ is  {\em dual Horn\/} if it is the set of
solutions of a dual Horn formula, where a dual Horn formula is a
\cnf~formula such that each conjunct has at most one negative
literal.
\item $R$ is {\em affine\/} if it is the set of
solutions of a   system of linear equations over $\mathbb{Z}_2$.
\end{enumerate}
%}
\end{definition}

Each of these types of logical relations can be characterized in
terms of {\em closure\/} properties \cite{Sc78}. A relation $R$ is
bijunctive if and only if it is closed under the {\em majority\/}
operation; this means that if $\mb{a}, \mb{b}, \mb{c} \in R$, then
$\maj(\mb{a}, \mb{b}, \mb{c}) \in R$, where $\maj(\mb{a}, \mb{b},
\mb{c})$ is the vector whose $i$-th bit is the majority of $a_i,
b_i, c_i$. A relation $R$ is Horn if and only if it is closed
under $\vee$; this means that if $\mb{a}, \mb{b} \in R$, then
$\mb{a}\vee \mb{b} \in R$, where, $\mb{a} \vee \mb{b}$ is the
vector whose $i$-th bit is $a_i\vee b_i$. Similarly, $R$ is dual
Horn if and only if it is closed under $\wedge$. Finally, $R$ is
affine if and only if it is closed under $\mb{a}\oplus
\mb{b}\oplus \mb{c}$. Thus there is a polynomial-time algorithm
%phk added
(in fact, a cubic algorithm) to test if a relation is Schaefer.

%  For instance,
%given $\mb{a}, \mb{b}, \mb{c} \in \{0,1\}^k$, the
%\emph{bitwise-majority}
% $\maj(\mb{a}, \mb{b}, \mb{c})$ is the vector
%whose $i$-th bit is the majority  of $a_i, b_i, c_i$. It is known
%(see \cite{Sc78}) that  a relation $R$ is bijunctive if and only
%iff it is \emph{closed} under majority, which means that if
%$\mb{a}, \mb{b}, \mb{c} \in R$, then $\maj(\mb{a}, \mb{b}, \mb{c})
%\in R$.  Similarly, define the \emph{bit-wise OR} of $\mb{a}$ and
%$\mb{b}$ to be the vector whose $i^{th}$ bit is $a_i \vee b_i$. A
%relation $R$ is Horn iff it is closed under bitwise OR and dual
%Horn iff it is closed under bitwise AND.

\begin{definition} \label{Schaefer-set} 
%{\rm
A set ${\cal S}$ of logical relations is {\em Schaefer\/} if at
least one of the following conditions holds:
\begin{enumerate}
\item Every relation in ${\cal S}$ is bijunctive.
\item Every relation in ${\cal S}$ is Horn.
\item Every relation in ${\cal S}$ is dual Horn.
\item Every relation in ${\cal S}$ is affine.
\end{enumerate}
%}
\end{definition}

\begin{theorem}
\label{thm:sch-dich} {\rm {(Schaefer's Dichotomy Theorem \cite{Sc78})}} 
Let $\cal S$ be a finite set of logical relations.
If ${\cal S}$ is Schaefer, then \sat$({\cal S})$ is in \PTIME;
otherwise, \sat$({\cal S})$ is \NP-complete.
\end{theorem}

%phk added Ladner's theorem discussion
Theorem \ref{thm:sch-dich} is called a dichotomy theorem because
Ladner \cite{La75} has shown that if ${\rm P} \not = {\rm NP}$,
then there are problems in NP that are neither in P, nor
NP-complete. Thus, Theorem \ref{thm:sch-dich} asserts that no
\sat$({\cal S})$ problem is  a problem of the kind discovered by
Ladner. Note that the aforementioned characterization of Schaefer
sets in terms of closure properties yields a cubic algorithm for
determining, given a finite set $\cal S$ of logical relations,
whether \sat$({\cal S})$ is in P or is NP-complete (here, the
input size is the sum of the sizes of the relations in $\cal S$).

 %phk rephrased
 %The
%hard side of Schaefer's dichotomy theorem
The more difficult part of the proof of Schafer's Dichotomy
Theorem is to show that if $\cal S$ is not Schaefer, then
\sat$({\cal S})$ is NP-complete. This  is a consequence of a
powerful result about the expressibility of logical relations. We
say that a relation $R$ is \emph{expressible from} a set ${\cal
S}$ of relations if there is a \cnf$({\cal S})$-formula
$\varphi({\bf x}, {\bf y})$ such that  $R= \{{\bf a}| \exists
\mb{y}~\varphi(\mb{a}, \mb{y})\}$.

\begin{theorem}
\label{thm:sch-expr} {\rm {(Schaefer's Expressibility Theorem
\cite{Sc78})}} Let $\cal S$ be a finite set of logical relations.
If ${\cal S}$ is not Schaefer, then every logical relation is
expressible from ${\cal S}$.
\end{theorem}

%Note that the closure properties of Schaefer sets yield a cubic
%algorithm for determining, given a finite set ${\cal S}$ of relations,
%whether \sat$({\cal S})$ is in \PTIME~or \NP-complete (the input size is the
%sum of the sizes of relations in ${\cal S}$).

In this paper, we are interested in the connectivity properties of
the space of solutions of \cnf$({\cal S})$-formulas. If $\varphi$
is a \cnf$({\cal S})$-formula with $n$ variables, then the
\emph{solution graph} $G(\varphi)$ of $\varphi$ denotes the
subgraph of the $n$-dimensional hypercube induced by the solutions
of $\varphi$. This means that the vertices of $G(\varphi)$ are the
solutions of $\varphi$, and there is an edge between two solutions
of $G(\varphi)$ precisely when they differ in exactly one
variable.

We consider the following two algorithmic problems for
\cnf$({\cal S})$-formulas.
\begin{Prob}{\rm
The {\em Connectivity Problem\/}  \conn$({\cal S})$:

Given a
 \cnf$({\cal S})$-formula $\varphi$, is $G(\varphi)$ connected?}
 \end{Prob}

 \begin{Prob} {\rm The {\em st-Connectivity Problem\/}  \stconn$({\cal S})$:

  Given a
\cnf$({\cal S})$-formula $\varphi$ and two solutions $\mb{s}$ and
$\mb{t}$ of $\varphi$, is there a path from $\mb{s}$ to $\mb{t}$
in $G(\varphi)$?}
\end{Prob}

%\begin{enumerate}
%\item  The {\em connectivity\/} problem
%\conn$({\cal S})$: given a
% \cnf$({\cal S})$-formula $\varphi$, is $G(\varphi)$ connected?
%\item The {\em st-connectivity\/} problem \stconn$({\cal S})$:
%given a \cnf$({\cal S})$-formula $\varphi$ and two solutions $\mb{s}$ and
%$\mb{t}$ of $\varphi$, is there a path from $\mb{s}$ to $\mb{t}$
%in $G(\varphi)$?
%\end{enumerate}

To pinpoint the computational complexity of \conn$({\cal S})$ and
\stconn$({\cal S})$, we need to introduce certain new types of
relations.
%that will be used to define the key concept of a \emph{tight} set
%of relations.

\begin{definition}
\label{def:comp-bijunctive} %{\rm  
Let $R\subseteq \{0,1\}^k$  be a logical relation.
\begin{enumerate}
\item $R$ is \emph{componentwise bijunctive} if every connected
component of the graph $G(R)$  is a bijunctive relation. 
\item $R$ is
\emph{\OR-free} if the  relation $\OR=\{01, 10, 11\}$ cannot be
obtained from $R$ by setting $k-2$ of the coordinates of $R$ to  a
constant $\mb{c}\in \{0,1\}^{k-2}$. In other words, $R$ is
\OR-free if $(x_1\lor x_2)$ is not definable from $R$ by fixing
$k-2$ variables. \item  $R$ is {\em \NAND-free\/} if the relation
$\NAND =\{00, 01,10\}$ cannot be obtained from $R$ by setting
$k-2$ of the coordinates of $R$ to a constant $\mb{c}\in
\{0,1\}^{k-2}$. In other words, $R$ is \NAND-free is
$(\bar{x}_1\vee \bar{x}_2)$ is not definable from $R$ by fixing
$k-2$ variables.
\end{enumerate}
%}
\end{definition}

We are now ready to introduce the key concept of a \emph{tight}
set of relations.

\begin{definition} %{\rm
A set ${\cal S}$ of logical relations is {\em tight\/} if at least one of
the following three conditions holds:
\begin{enumerate}
\item Every relation in ${\cal S}$ is componentwise bijunctive;
\item Every relation in ${\cal S}$ is \OR-free;
\item Every relation in ${\cal S}$ is \NAND-free.
% Otherwise, we say that ${\cal S}$ is {\em not tight}.
\end{enumerate}
%}
\end{definition}

In Section \ref{sec:easy}, we show that if ${\cal S}$ is Schaefer,
then it is tight. Moreover, we show that the converse does not
hold. It is also easy to see that there is a polynomial-time
algorithm
%phk added cubic
(in fact, a cubic algorithm)
for testing whether a given
relation is tight.

%phk edited
Just as Schaefer's dichotomy theorem follows from an
expressibility statement, our dichotomy theorems are derived from
the following theorem, which we will call the \FET.
%At
%a high level, this theorem asserts that any logical relation with
%a solution graph $G$ is expressible from any non-tight set of
%relations, in such a way that the solution graph of the formula is
%isomorphic to $G$ after certain adjacent vertices are merged.
The precise definition of the concept of \emph{faithful
expressibility} is given in Section \ref{sec:hard}. Intuitively,
this concept strengthens the concept of expressibility with the
requirement that the space of the witnesses to the existentially
quantified variables has certain strong connectivity properties.

\begin{theorem} {\rm(\FET)}
\label{thm:expr} Let $\cal S$ be a finite set of logical
relations. If ${\cal S}$  is not tight, then every  logical
relation is faithfully expressible from ${\cal S}$.
\end{theorem}

Using the \FET, we obtain the following dichotomy theorems for the
computational complexity of \conn$({\cal S})$ and \stconn$({\cal
S})$.

\begin{theorem}
\label{thm:conn} Let ${\cal S}$ be a finite set of logical relations.
If ${\cal S}$ is tight, then \conn$({\cal S})$ is in \coNP; otherwise, it
is \PSPACE-complete.
\end{theorem}

\begin{theorem}
\label{thm:stconn} Let ${\cal S}$ be a finite set of logical relations.
If ${\cal S}$ is tight, then \stconn$({\cal S})$ is in \PTIME; otherwise,
\stconn$({\cal S})$ is \PSPACE-complete.
\end{theorem}

We also show that if ${\cal S}$ is tight, but not Schaefer, then
\conn$({\cal S})$ is \coNP-complete.

To illustrate these results, consider   the set ${\cal
S}=\{R_{1/3}\}$, where $R_{1/3}=\{100, 010, 001\}$. This set  is
tight (actually, it is componentwise bijunctive), but not
Schaefer. It follows that \sat$({\cal S})$ is \NP-complete (recall
that this problem is {\sc Positive 1-in-3 \sat}), \stconn$({\cal
S})$ is in \PTIME, and \conn$({\cal S})$ is \coNP-complete.
Consider also the set ${\cal S}=\{R_{\rm {NAE}}\}$, where $R_{\rm
{NAE}}=\{0,1\}^3\setminus \{000, 111\}$. This set is not tight,
hence \sat$({\cal S})$ is \NP-complete (this problem is {\sc
Positive Not-All-Equal 3-\sat}), while both \stconn$({\cal S})$
and \conn$({\cal S})$ are \PSPACE-complete.

The dichotomy in the computational complexity of \conn$({\cal S})$ and
\stconn$({\cal S})$ is accompanied by a parallel structural dichotomy
in the size of the diameter of $G(\varphi)$ (where, for a \cnf$({\cal
S})$-formula $\varphi$, the
\emph{diameter of $G(\varphi)$} is the maximum of the diameters of
the components of $G(\varphi)$).

\begin{theorem}
\label{thm:diam} Let ${\cal S}$ be a finite set of logical relations.
If ${\cal S}$ is tight, then for every \cnf$({\cal S})$-formula $\varphi$,
the diameter of $G(\varphi)$ is linear in
the number of variables of $\varphi$;
otherwise, there are   \cnf$({\cal S})$-formulas $\varphi$ such that
the diameter of $G(\varphi)$ is exponential in the number of
variables of $\varphi$.
\end{theorem}

Our results and their comparison to Schaefer's Dichotomy Theorem are
summarized in the table below.

\smallskip
\begin{center}
\begin{tabular}{|l|l|l|l|l|}
\hline
${\cal S}$ & \sat$({\cal S})$ &  \stconn$({\cal S})$ & \conn$({\cal S})$  & Diameter\\
\hline
Schaefer & \PTIME & \PTIME & \coNP & $O(n)$\\
Tight, non-Schaefer & \NP-compl. & \PTIME & \coNP-compl.  & $O(n)$\\
Non-tight & \NP-compl. & \PSPACE-compl. & \PSPACE-compl. &
$2^{\Omega(\sqrt{n})}$ \\
\hline
\end{tabular}
\end{center}
\vspace{0.5cm}

We conjecture that the complexity of \conn$({\cal S})$ exhibits a
\emph{trichotomy}, that is, for every finite set $\cal S$ of
logical relations, one of the following holds:
\begin{enumerate}
\item \conn$({\cal S})$ is in \PTIME; \item  \conn$({\cal S})$ is
coNP-complete; \item \conn$({\cal S})$ is PSPACE-complete.
\end{enumerate}

As mentioned above,  we will show that if ${\cal S}$ is tight but
not Schaefer, then \conn$({\cal S})$ is \coNP-complete. We will
also show that if ${\cal S}$ is bijunctive or affine, then
\conn$({\cal S})$ is in \PTIME. Hence, to settle the above
conjecture, it remains to pinpoint the complexity of \conn$({\cal
S})$ whenever $\cal S$ is  Horn   and whenever $\cal S$ is  dual
Horn. In the conference version \cite{icalp} of the present paper,
we further conjectured that if ${\cal S}$ is Horn or dual Horn,
then \conn$({\cal S})$ is in \PTIME. In other words, we
conjectured that if ${\cal S}$ is Schaefer, then \conn$({\cal S})$
is in \PTIME.
 This second conjecture, however,  was subsequently disproved  by
 Makino, Tanaka and Yamamato \cite{MTY07}, who discovered a particular Horn set
${\cal S}$ such that \conn$({\cal S})$ is \coNP-complete. Here, we
go beyond the results obtained in the conference version of the
present paper and identify additional conditions on a Horn set
$\cal S$ implying that \conn$({\cal S})$ is in \PTIME. These new
results suggest a natural dichotomy within Schaefer sets of
relations and, thus, provide evidence for the trichotomy
conjecture.

The remainder of this paper is organized as follows. In Section
\ref{sec:hard},  we prove the Faithful Expressibility Theorem,
establish the hard side of the dichotomies for \conn$({\cal S})$
and for \stconn$({\cal S})$,  and   contrast our result to
Schaefer's Expressibility and Dichotomy Theorems. In Section
\ref{sec:easy}, we describe the easy side of the dichotomy - the
polynomial-time algorithms and the structural properties for tight
sets of relations. In addition,
 we obtain partial results
towards the trichotomy conjecture for \conn$({\cal S})$.

\section{The Hard Case of the Dichotomy: Non-Tight Sets of Relations}
\label{sec:hard}

%phk
%We will show that all non-tight sets of relations lead to solution
%graphs that have identical properties in a natural sense that is
%captured in the notion of faithful expressibility.

%\pnote{Is this ok for the hard issue?}
In this section, we address the {\em hard} side of the dichotomy,
where we deal with the more computationally intractable cases. As with
other dichotomy theorems, this is also the harder part of our proof.
We define the notion of faithful expressibility in Section
\ref{sec:expr} and prove the Faithful Expressibility Theorem in
Section \ref{sec:FET}. This theorem implies that for all
non-tight sets $\cal S$ and ${\cal S'}$,   the connectivity
problems \conn$({\cal S})$ and \conn$({\cal S'})$ are
polynomial-time equivalent; moreover, the same holds true for the
connectivity problems \stconn$({\cal S})$ and \stconn$({\cal
S'})$.  In addition, the diameters of the solution graphs of
\cnf$({\cal S})$-formulas and \cnf$({\cal S'})$-formulas are also
related polynomially. In Section \ref{sec:3sat}, we prove that for
3-\cnf~formulas the connectivity problems are \PSPACE-complete,
and the diameter can be exponential. This fact combined with the
Faithful Expressibility Theorem yields the hard side of all of our
dichotomy results, as well as the exponential  size of the
diameter.

We will use $\mb{a}, \mb{b},\dots $ to denote Boolean vectors, and
$\mb{x}$ and $\mb{y}$ to denote vectors of variables.  We write
$|\mb{a}|$ to denote the Hamming weight (number of $1$'s) of a Boolean
vector $\mb{a}$. Given two Boolean vectors $\mb{a}$ and $\mb{b}$, we
write $|\mb{a} - \mb{b}|$ to denote the Hamming distance between
$\mb{a}$ and $\mb{b}$.  Finally, if $\mb{a}$ and $\mb{b}$ are
solutions of a Boolean formula $\varphi$ and lie in the same component
of $G(\varphi)$, then we write $d_\varphi(\mb{a}, \mb{b})$
% (or, $d(\mb{a},\mb{b})$
%when $\varphi$ is understood from the context)
to denote the shortest-path distance between $\mb{a}$ and $\mb{b}$ in
$G(\varphi)$.

\subsection{Faithful Expressibility}
\label{sec:expr}

As we mentioned in the previous section, in
his dichotomy theorem, Schaefer \cite{Sc78} used the following
notion of expressibility: a relation $R$ is \emph{expressible from} a
set ${\cal S}$ of relations if there is a
\cnf$({\cal S})$-formula $\varphi$ so that 
%$R(\mb{x}) \equiv \exists \mb{y}~\varphi(\mb{x}, \mb{y})$. 
$R = \{\mb{a}|~ \exists \mb{y}~\varphi(\mb{a}, \mb{y})\}$.
This notion, is not
sufficient for our purposes. Instead, we introduce a more delicate
notion, which we call  {\em faithful expressibility}. Intuitively, we
view the relation $R$ as a subgraph of the hypercube, rather than
just a subset, and require that this graph structure be also
captured by the formula $\varphi$.

\begin{figure}
\label{fig:expr}
\begin{center}
\begin{tabular}{c}
\epsfig{file=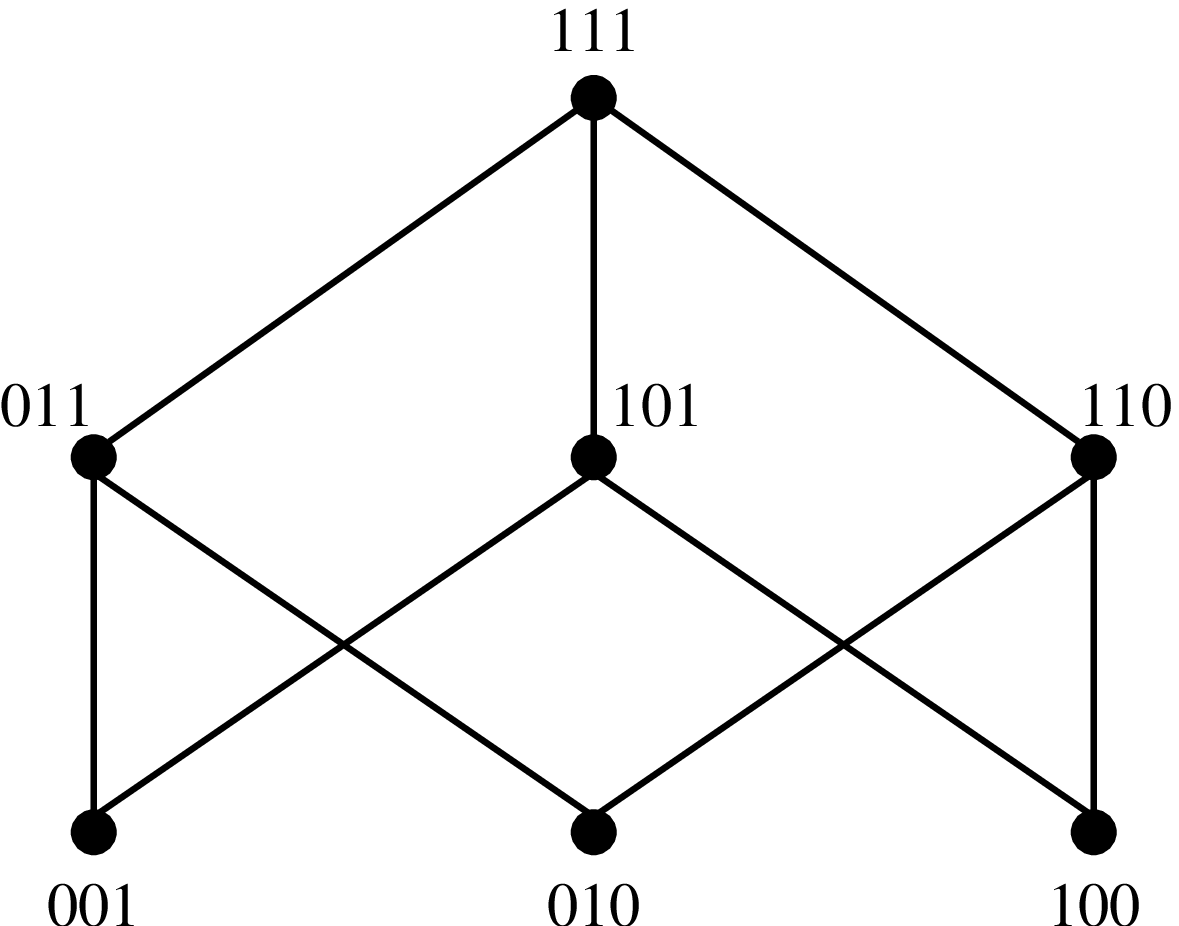, height=3.7cm}\\
(a)
\\ \\
\epsfig{file=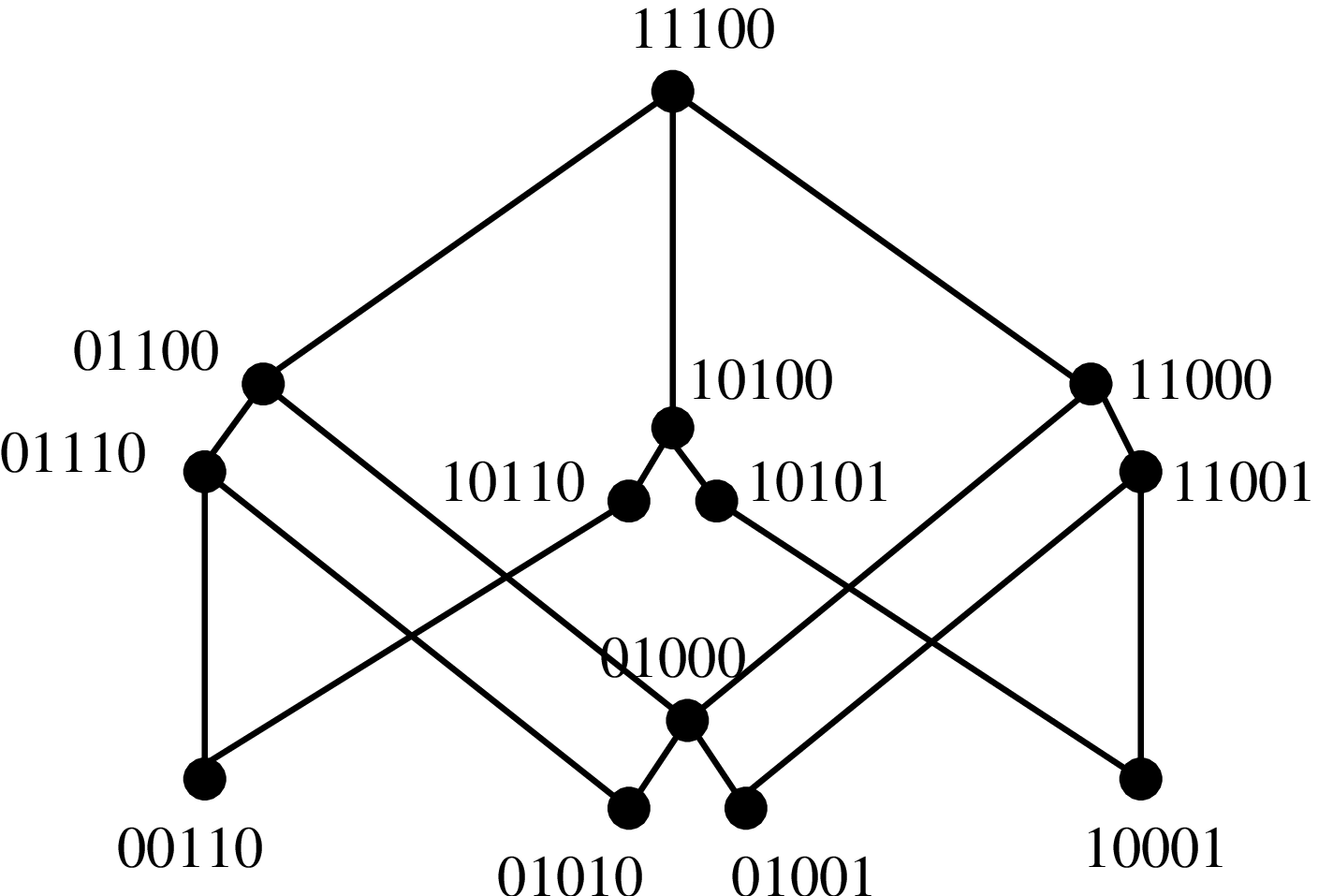, height=3.7cm}
~~~~
\epsfig{file=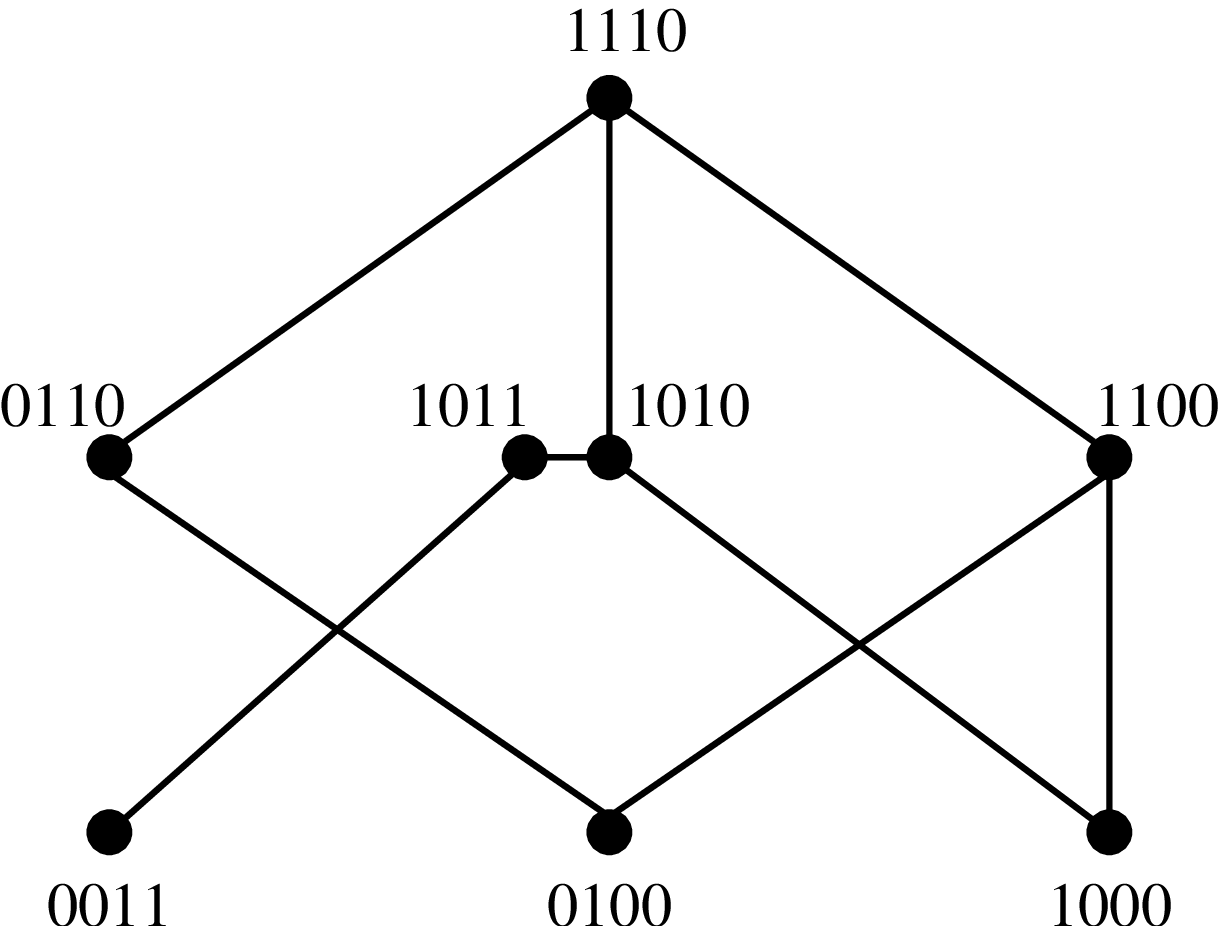, height=3.7cm} \\  \\
(b) ~~~~~~~~~~~~~~~~~~~~~~~~~~~~~~~~~~~~~~~~~~~~~~~~~~~~~~~(c)
\end{tabular}
\caption{Expressing the relation $(x_1 \vee x_2 \vee x_3)$
using {\sc Not-All-Equal} relations. \newline
(a) Graph of $(x_1 \vee x_2 \vee x_3)$; \newline
(b) Graph of a faithful expression: $\varphi(\mb{x}, y_1, y_2) =
R_{\rm {NAE}}(x_1, x_2, y_1) \wedge R_{\rm {NAE}}(x_2, x_3, y_2) \wedge R_{\rm {NAE}}(y_1, y_2, 1).$  \newline
(c) Graph of an unfaithful expression:
$\varphi(\mb{x}, y_1) = R_{\rm {NAE}}(x_1, x_2, y_1) \wedge R_{\rm {NAE}}(\bar{y}_1, x_3, 0) \wedge R_{\rm {NAE}}(y_1, x_2, 1).$  \newline
In both cases $(x_1 \vee x_2 \vee x_3) = \exists \mb{y}~
\varphi(\mb{x}, \mb{y})$, but only in the first case the
connectivity is preserved.}
\end{center}
\end{figure}

\begin{definition}
\label{def:connected} A relation $R$ is {\em faithfully expressible}
from a set of relations ${\cal S}$ if there is a \cnf$({\cal S})$-formula
$\varphi$ such that the following conditions hold:
\begin{enumerate}
\item $R = \{\mb{a}|~ \exists \mb{y}~\varphi(\mb{a}, \mb{y})\}$;
\item For every $\mb{a} \in R$, the graph $G(\varphi(\mb{a},
\mb{y}))$ is connected;
\item For $\mb{a}, \mb{b} \in R$ with $|\mb{a}-\mb{b}|=1$, there
exists $\mb{w}$ such that $(\mb{a},\mb{w})$
and $(\mb{b},\mb{w})$ are solutions of $\varphi$.
\end{enumerate}
\end{definition}

For $\mb{a} \in R$, the {\em witnesses\/} of $\mb{a}$ are the
$\mb{y}$'s such that $\varphi(\mb{a}, \mb{y})$ is true.  The last two
conditions say that the witnesses of $\mb{a}\in R$ are connected, and
that neighboring $\mb{a}, \mb{b} \in R$ have a common witness.  This
allows us to simulate an edge $(\mb{a}, \mb{b})$ in $G(R)$ by a path
in $G(\varphi)$, and thus relate the connectivity properties of the
solution spaces. There is however, a price to pay: it is much harder
to come up with formulas that faithfully express a relation $R$. An
example is when ${\cal S}$ is the set of all paths of length $4$ in
$\{0,1\}^3$, a set that plays a crucial role in our proof. While
3-\sat~relations are easily expressible from ${\cal S}$ in Schaefer's sense,
the \cnf$({\cal S})$-formulas that faithfully express 3-\sat~relations are
fairly complicated and have a large witness space.

An example of the difference between a faithful and an unfaithful
expression is shown in Figure \ref{fig:expr}.

\begin{lemma}
\label{lem:red} Let ${\cal S}$ and ${\cal S}'$ be sets of relations such that
every $R \in {\cal S}'$ is faithfully expressible from ${\cal S}$. Given a
\cnf$({\cal S}')$-formula $\psi(\mb{x})$,  one can efficiently construct a
\cnf$({\cal S})$-formula $\varphi(\mb{x}, \mb{y})$ such that:
\begin{enumerate}
\item $\psi(\mb{x})\equiv \exists \mb{y}~\varphi(\mb{x}, \mb{y})$;
\item if $(\mb{s}, \mb{w^s}), (\mb{t}, \mb{w^t}) \in \varphi$ are
connected in $G(\varphi)$ by a path of length $d$, then there is a
path from $\mb{s}$ to $\mb{t}$ in $G(\psi)$ of length at most $d$;
\item If $\mb{s}, \mb{t} \in \psi$ are connected in $G(\psi)$,
then for every witness $\mb{w^s}$ of $\mb{s}$, and every witness
$\mb{w^t}$ of $\mb{t}$, there is a path from $(\mb{s}, \mb{w^s})$
to $(\mb{t}, \mb{w^t})$ in $G(\varphi)$.
\end{enumerate}
\end{lemma}
\begin{proof}
Suppose $\psi$ is a formula on $n$ variables that consists of $m$
clauses $C_1, \dots, C_m$. For clause $C_j$, assume that the
set of variables is $V_j \subseteq [n]$,
and that it involves relation $R_j \in {\cal S}$. Thus,
$\psi(\mb{x})$ is $\wedge_{j=1}^m R_j(\mb{x}_{V_j})$. Let
$\varphi_j$ be the faithful expression for $R_j$ from ${\cal S}'$, so that
$R_j(\mb{x}_{V_j})\equiv \exists \mb{y}_j~\varphi_j(\mb{x}_{V_j},
\mb{y}_j)$. Let $\mb{y}$ be the vector $(\mb{y}_1,\dots, \mb{y}_m)$
and let $\varphi(\mb{x},\mb{y})$ be the formula
$\wedge_{j=1}^m\varphi_j(\mb{x}_{V_j},\mb{y}_j)$. Then
$\psi(\mb{x}) \equiv \exists \mb{y}~\varphi(\mb{x},\mb{y})$.

Statement $(2)$ follows from $(1)$ by projection of the path on the
coordinates of $\mb{x}$.  For statement $(3)$, consider
$\mb{s}, \mb{t} \in \psi$ that are connected in $G(\psi)$ via a path
$\mb{s}=\mb{u^0} \rightarrow \mb{u^1} \rightarrow
\dots \rightarrow \mb{u^r}=\mb{t}$ . For every $\mb{u^i},
\mb{u^{i+1}}$, and clause $C_j$, there exists an assignment
$\mb{w^i}_j$ to $\mb{y}_j$ such that both $(\mb{u^i}_{V_j},
\mb{w^i}_j)$ and $(\mb{u^{i+1}}_{V_j}, \mb{w^i}_j)$ are solutions of
$\varphi_j$, by condition $(2)$ of faithful expressibility. Thus
$(\mb{u^i}, \mb{w^i})$ and $(\mb{u^{i+1}}, \mb{w^i})$ are both
solutions of $\varphi$, where $\mb{w^i}=(\mb{w^i}_1, \dots
\mb{w^i}_m)$.  Further, for every $\mb{u^i}$, the space of solutions
of $\varphi(\mb{u^i}, \mb{y})$ is the product space of the solutions
of $\varphi_j(\mb{u^i}_{V_j},
\mb{y}_j)$ over $j=1, \dots, m$. Since these are all connected by
condition $(3)$ of faithful expressibility, $G(\varphi(\mb{u^i},
\mb{y}))$ is connected.  The following describes a path from
$(\mb{s}, \mb{w^s})$ to $(\mb{t}, \mb{w^t})$ in $G(\varphi)$:
~$(\mb{s}, \mb{w^s}) \rightsquigarrow (\mb{s},
\mb{w^0})\rightarrow (\mb{u^1}, \mb{w^0}) \rightsquigarrow
(\mb{u^1}, \mb{w^1}) \rightarrow \dots \rightsquigarrow
(\mb{u^{r-1}}, \mb{w^{r-1}}) \rightarrow (\mb{t}, \mb{w^{r-1}})
\rightsquigarrow (\mb{t}, \mb{w^t})$. Here $\rightsquigarrow$
indicates a path in $G(\varphi(\mb{u^i}, \mb{y}))$.
\end{proof}

\begin{corollary}
\label{cor:red} Suppose ${\cal S}$ and ${\cal S}'$
are sets of relations such that every $R \in {\cal S}'$ is faithfully
expressible from ${\cal S}$.
\begin{enumerate}
\item There are polynomial time reductions from
\conn$({\cal S'})$ to \conn$({\cal S})$, and from \stconn$({\cal S'})$ to
\stconn$({\cal S})$.
\item Given a \cnf$({\cal S'})$-formula $\psi(\mb{x})$ with $m$
clauses, one can efficiently construct a \cnf$({\cal S})$-formula
$\varphi(\mb{x}, \mb{y})$ such that  the length of $\mb{y}$ is
$O(m)$ and the diameter of the solution space does not decrease.
\end{enumerate}
\end{corollary}

%In the next section we will prove that the conditions for the above
%corollary hold whenever ${\cal S}$ is not a tight set of relations,
%which is the statement of Theorem \ref{thm:expr}.

\subsection{The \FET}
\label{sec:FET}

In this subsection, we prove the \FET.
The main step in the proof is Lemma \ref{lem:expr} which
shows that if ${\cal S}$ is not tight, then  we can faithfully express
the 3-clause relations from the relations in ${\cal S}$.
If $k\geq 2$, then a $k$-{\em clause\/} is a disjunction of $k$ variables or negated
variables. For $0 \leq i \leq k$, let $D_i$ be the set of all
satisfying truth assignments of the $k$-clause whose first $i$
literals are negated, and let ${\cal S}_k=\{D_0,D_1,\dots,D_k\}$. Thus,
\cnf$({\cal S}_k)$ is the collection of $k$-\cnf~formulas.

\begin{lemma}
\label{lem:expr} If set ${\cal S}$ of relations is not tight,
${\cal S}_3$ is faithfully expressible from ${\cal S}$.
\end{lemma}

\begin{proof}
First, observe that all $2$-clauses are faithfully expressible from
${\cal S}$.  There exists $R \in {\cal S}$ which is not \OR-free, so
we can express $(x_1 \orr x_2)$ by substituting constants in
$R$. Similarly, we can express $(\bar{x}_1 \vee \bar{x}_2)$ using a
relation that is not \NAND-free. The last 2-clause $(x_1 \orr
\bar{x}_2)$ can be obtained from \OR~and \NAND~by a technique that
corresponds to reverse resolution.  $ (x_1 \orr \bar{x}_2) = \exists
y~(x_1 \orr y) \ann (\bar{y} \orr \bar{x}_2)$. It is easy to see that
this gives a faithful expression.  From here onwards we assume that
${\cal S}$ contains all 2-clauses.  The proof now proceeds in four
steps. First, we will express a relation in which there exist two
elements that are at graph distance larger than their Hamming
distance.  Second, we will express a relation that is just a single
path between such elements. Third, we will express a relation which
is a path of length 4 between elements at Hamming distance 2. Finally,
we will express the 3-clauses.

\begin{step}
\label{step:1}
Faithfully expressing a relation in which some distance expands.
\end{step}
For a relation $R$, we say that the distance between $\mb{a}$ and
$\mb{b}$ {\em expands} if $\mb{a}$ and $\mb{b}$ are connected in
$G(R)$, but $ d_R(\mb{a}, \mb{b}) > |\mb{a} - \mb{b}|$.
%phk
Later on, we will show that
 no
distance expands in componentwise bijunctive relations. The same
also holds true for the relation $R_{\rm {NAE}}=\{0,1\}^3\setminus
\{000, 111\}$, which is not componentwise bijunctive. Nonetheless,
we show here that if $R$ is not componentwise bijunctive, then, by
adding $2$-clauses, we can faithfully express a relation $Q$ in
which some distance expands. For instance, when $R = R_{\rm NAE}$,
then we can take $Q(x_1,x_2,x_3) = R_{\rm NAE}(x_1,x_2,x_3) \wedge
(\bar{x_1} \vee \bar{x}_3)$. The distance between $\mb{a} = 100$
and $\mb{b} = 001$ in $Q$ expands. Similarly, in the general
construction, we identify $\mb{a}$ and $\mb{b}$ on a cycle, and
add $2$-clauses that eliminate all the vertices along the shorter
arc between $\mb{a}$ and $\mb{b}$.

\begin{figure}
\label{fig:expand}
\begin{center}
\begin{tabular}{c}
\epsfig{file=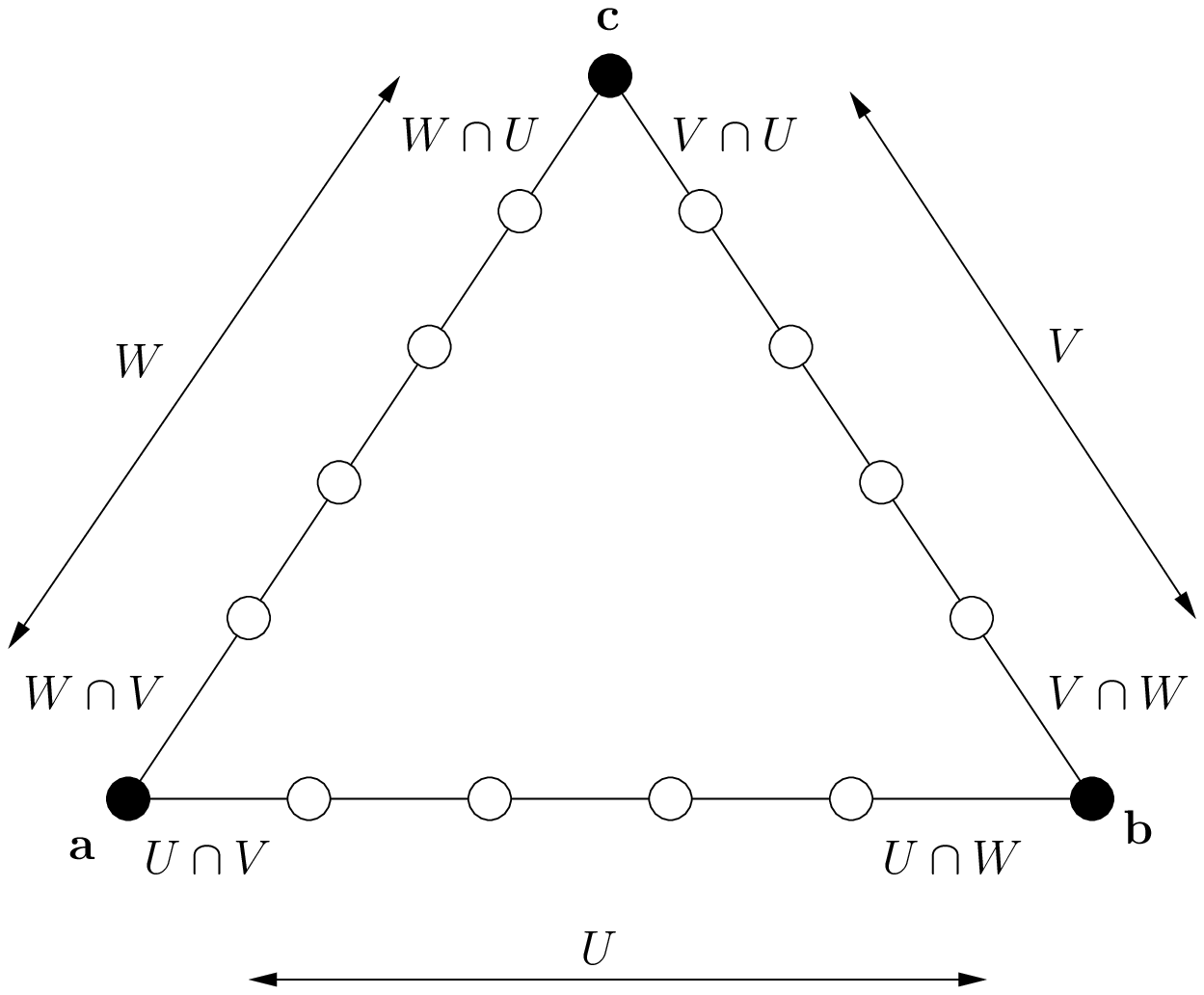, height=4.7cm}
\\ \\
\epsfig{file=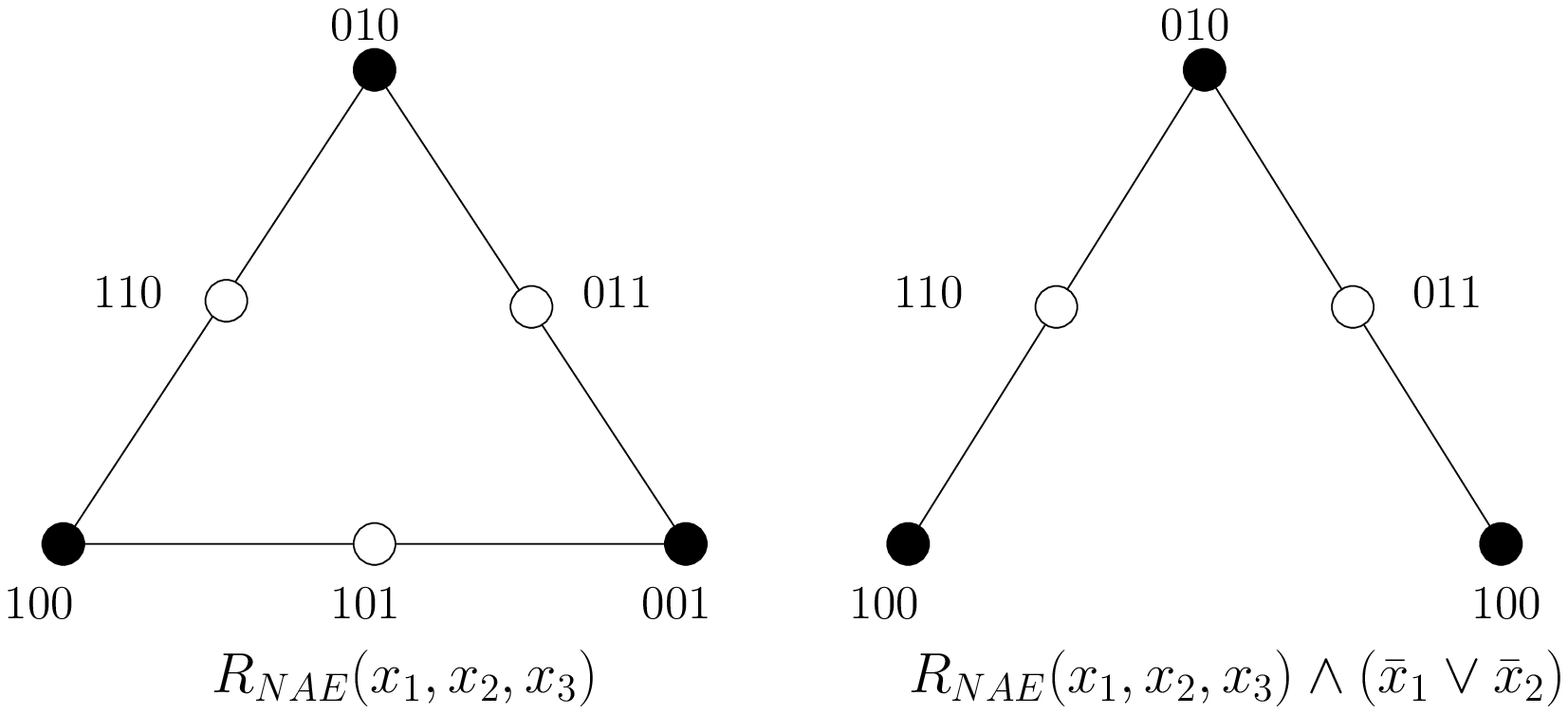, height=4.7cm}
\end{tabular}
\caption{Proof of Step \ref{step:1} of Lemma \ref{lem:expr}, and an example.}
\end{center}
\end{figure}

\comment{
\begin{lemma}
\label{lem:expand} There exist a \cnf$({\cal S})$-definable relation $Q$
 and $\mb{a}, \mb{b} \in Q$ such that the distance between them
%in $Q$
expands.
\end{lemma}
\begin{proof}
}

Since ${\cal S}$ is not tight, it contains a relation $R$ which is not
componentwise bijunctive.  If $R$ contains $\mb{a}, \mb{b}$ where
the distance between them expands, we are done. So assume that for all
$\mb{a}, \mb{b} \in G(R)$, $ d_R(\mb{a},\mb{b}) = |\mb{a} -
\mb{b}|$.  Since $R$ is not componentwise bijunctive, there exists a
triple of assignments $\mb{a}, \mb{b}, \mb{c}$ lying in the same
component such that $\maj(\mb{a}, \mb{b}, \mb{c})$ is not in that component
(which also easily implies it is not in $R$).
Choose the triple such that the sum of pairwise
distances $d_R(\mb{a}, \mb{b}) + d_R(\mb{b}, \mb{c}) + d_R(\mb{c}, \mb{a})$
is minimized. Let $U=\{ i| a_i\neq b_i\}$, $V=\{i|b_i\neq c_i\}$, and
$W = \{i|c_i\neq a_i\}$. Since $d_R(\mb{a}, \mb{b}) = |\mb{a} -
\mb{b}|$, a shortest path does not flip variables outside of $U$,
and each variable in $U$ is flipped exactly once. The same holds for $V$
and $W$. We note some useful properties of the sets $U, V, W$.

\begin{enumerate}
\item {\em Every index $i \in U \cup V \cup W$ occurs in exactly two of
$U, V, W$.}\\
Consider going by a shortest path from $\mb{a}$ to
$\mb{b}$ to $\mb{c}$ and back to $\mb{a}$. Every $i \in U \cup V
\cup W$ is seen an even number of times along this path since we
return to $\mb{a}$. It is seen at least once, and at most thrice,
so in fact it occurs twice.

\item {\em Every pairwise intersection $U \cap V, V\cap W$ and $W \cap U$
is non-empty.}\\
Suppose the sets $U$ and $V$ are disjoint. From
Property 1, we must have $W = U \cup V$.
But then it is easy to see that $\maj(\mb{a}, \mb{b}, \mb{c}) = \mb{b}$
which is in $R$.
This contradicts the choice of $\mb{a}, \mb{b}, \mb{c}$.

\item {\em The sets $U \cap V$ and $U \cap W$ partition the set $U$}.\\
By Property $1$, each index of $U$ occurs in one of $V$ and $W$ as well. Also
since no index occurs in all three sets $U, V, W$  this is
in fact a disjoint partition.

\item {\em For each index $i \in U \cap W$, it holds that $\mb{a} \oplus
\mb{e}_i \not\in R$.}\\
Assume for the sake of contradiction that
$\mb{a'} = \mb{a} \oplus \mb{e}_i \in R$. Since $i \in U \cap W$ we
have simultaneously moved closer to both $\mb{b}$ and $\mb{c}$. Hence
we have $ d_R(\mb{a'}, \mb{b}) + d_R(\mb{b}, \mb{c}) + d_R(\mb{c},
\mb{a'}) < d_R(\mb{a}, \mb{b}) + d_R(\mb{b}, \mb{c}) + d_R(\mb{c}, \mb{a})$.
Also $\maj(\mb{a'},\mb{b}, \mb{c}) = \maj(\mb{a}, \mb{b}, \mb{c})
\not\in R$. But this contradicts our choice of $\mb{a}, \mb{b},
\mb{c}$.
\end{enumerate}

Property 4 implies that the shortest paths to $\mb{b}$ and $\mb{c}$
diverge at $\mb{a}$, since for any shortest path to $\mb{b}$ the first
variable flipped is from $U \cap V$ whereas for a shortest path to
$\mb{c}$ it is from $W \cap V$.  Similar statements hold for the
vertices $\mb{b}$ and $\mb{c}$.  Thus along the shortest path from
$\mb{a}$ to $\mb{b}$ the first bit flipped is from $U \cap V$ and the
last bit flipped is from $U \cap W$. On the other hand, if we go from
$\mb{a}$ to $\mb{c}$ and then to $\mb{b}$, all the bits from $U \cap
W$ are flipped before the bits from $U \cap V$.  We use this
crucially to define $Q$. We will add a set of 2-clauses
that enforce the following rule on paths starting at $\mb{a}$:
{\em Flip variables from $U \cap W$ before variables from $U \cap V$.}
This will eliminate all shortest paths from $\mb{a}$ to
$\mb{b}$ since they begin by flipping a variable in $U \cap V$ and
end with $U \cap W$.  The paths from $\mb{a}$ to $\mb{b}$ via $\mb{c}$
survive since they flip $U \cap W$ while going from $\mb{a}$ to
$\mb{c}$ and $U \cap V$ while going from $\mb{c}$ to $\mb{b}$. However
all remaining paths have length at least $|\mb{a}-\mb{b}|+2$ since
they flip twice some variables not in $U$.

Take all pairs of indices $\{(i,j)| i \in U\cap W, j \in
U \cap V\}$. The following conditions hold from the definition of
$U,V,W$: $ a_i = \bar{c}_i = \bar{b}_i$ and $a_j = c_j = \bar{b}_j
$. Add the 2-clause $C_{ij}$ asserting that the pair of variables
$x_ix_j$ must take values in $\{a_ia_j, c_ic_j, b_ib_j \} =
\{a_ia_j, \bar{a}_ia_j, \bar{a}_i\bar{a}_j \}$.  The new relation
is $Q=R \ann_{i,j} C_{ij}$. Note that $Q \subset R$.
We verify that the distance between $\mb{a}$ and $\mb{b}$ in $Q$
expands. It is easy to see that for any $j \in U$, the assignment
$\mb{a} \oplus \mb{e}_j \not\in Q$.
Hence there are no shortest paths left from $\mb{a}$ to $\mb{b}$.
On the other hand, it is easy to see that $\mb{a}$ and $\mb{b}$ are
still connected, since the vertex $\mb{c}$ is still reachable from
both.
%\end{proof}

\begin{step}
\label{step:2}
Isolating a pair of assignments whose distance expands.
\end{step}
The relation $Q$ obtained in Step \ref{step:1} may have
several disconnected components. This {\em cleanup} step isolates a
single pair of assignments whose distance expands. By adding
$2$-clauses, we show that one can express a path of length $r+2$
between assignments at distance $r$.

\comment{
\begin{lemma}
\label{lem:q-to-path} There is a relation $T$
faithfully expressible from $S$ so that $G(T)$ is a path of length
$r +2$ between assignments at Hamming distance $r$, for
$r\geq2$.
\end{lemma}
\begin{proof}
}

Take $\mb{a}, \mb{b} \in Q$ whose distance expands in $Q$ and $d_{Q}(\mb{a},
\mb{b})$ is minimized. Let $U=\{i|a_i \neq b_i\}$, and $|U| = r$.
Shortest paths between $\mb{a}$ and $\mb{b}$ have certain
useful properties:

\begin{enumerate}
\item {\em Each shortest path flips every variable from $U$ exactly
  once.}\\
Observe
that each index $j \in U$ is flipped an odd number of times along any
path from $\mb{a}$ to $\mb{b}$.  Suppose it is flipped thrice along a
shortest path. Starting at $\mb{a}$ and going along this path, let
$\mb{b'}$ be the assignment reached after flipping $j$ twice. Then the
distance between $\mb{a}$ and $\mb{b'}$ expands, since $j$ is flipped
twice along a shortest path between them in $Q$. Also $d_{Q}(\mb{a},
\mb{b'}) < d_{Q}(\mb{a}, \mb{b})$, contradicting the choice of
$\mb{a}$ and $\mb{b}$.

\item {\em Every shortest path flips exactly one variable $i \not\in U$. }\\
Since the distance between $\mb{a}$ and $\mb{b}$ expands,
every shortest path must flip some variable $i \not\in U$. Suppose it
flips more than one such variable. Since $\mb{a}$ and $\mb{b}$ agree
on these variables, each of them is flipped an even number of times. Let
$i$ be the first variable to be flipped twice. Let $\mb{b'}$ be the assignment
reached after flipping $i$ the second time. It is easy to verify that
the distance between $\mb{a}$ and $\mb{b'}$ also expands, but
$d_{Q}(\mb{a}, \mb{b'}) < d_{Q}(\mb{a}, \mb{b})$.

\item {\em The variable $i \not\in U$ is the first and last variable to be
flipped along the path. }
Assume the first variable flipped is not $i$. Let
$\mb{a'}$ be the assignment reached along the path before we flip $i$
the first time. Then $d_{Q}(\mb{a'}, \mb{b}) < d_{Q}(\mb{a},
\mb{b})$. The distance between $\mb{a'}$ and $\mb{b}$ expands since
the shortest path between them flips the variables $i$ twice. This
contradicts the choice of $\mb{a}$ and $\mb{b}$.  Assume $j \in U$ is
flipped twice. Then as before we get a pair $\mb{a'},
\mb{b'}$ that contradict the choice of $\mb{a}, \mb{b}$.
\end{enumerate}

Every shortest path between $\mb{a}$ and $\mb{b}$ has the following
structure: first a variable $i \not\in U$ is flipped to $\bar{a}_i$,
then the variables from $U$ are flipped in some order, finally the
variable $i$ is flipped back to $a_i$.

Different shortest paths may vary in the choice of $i \not\in U$ in
the first step and in the order in which the variables from $U$ are flipped.
Fix one such path $T \subseteq Q$.  Assume that $U =\{1,
\dots, r\}$ and the variables are flipped in this order, and the
additional variable flipped twice is $r+1$.  Denote the path by $\mb{a}
\rightarrow \mb{u^0} \rightarrow \mb{u^1} \rightarrow
\dots \rightarrow \mb{u^r} \rightarrow \mb{b}$.
Next we prove that we cannot flip the $r+1^{th}$ variable at
an intermediate vertex along the path.

\begin{enumerate}
\item[4]  {\em For $1 \leq j \leq r-1$ the assignment $\mb{u^j} \oplus
\mb{e_{r+1}} \not\in Q$. } \\
Suppose that for some $j$, we have $\mb{c} = \mb{u^j} \oplus
\mb{e_{r+1}} \in Q$. Then $\mb{c}$ differs from $\mb{a}$ on $\{1,
\dots, i\}$ and from $\mb{b}$ on $\{i+1, \dots, r\}$. The distance
from $\mb{c}$ to at least one of $\mb{a}$ or $\mb{b}$ must expand,
else we get a path from $\mb{a}$ to $\mb{b}$ through $\mb{c}$ of
length $|\mb{a} - \mb{b}|$ which contradicts the fact that this
distance expands. However $d_{Q}(\mb{a}, \mb{c})$ and $d_{Q}(\mb{b},
\mb{c})$ are strictly less than $d_{Q}(\mb{a}, \mb{b})$ so we get a
contradiction to the choice of $\mb{a}, \mb{b}$.
\end{enumerate}

We now construct the path of length $r+2$.  For
all $i \geq r +2$ we set $x_i =a_i$ to get a relation on $r +1$
variables. Note that $\mb{b} = \bar{a}_1\dots
\bar{a}_ra_{r+1}$.  Take $i <j \in U$. Along the path $T$ the
variable $i$ is flipped before $j$ so the variables $x_ix_j$ take one of
three values $ \{a_ia_j, \bar{a}_ia_j, \bar{a}_i\bar{a}_j \}$.
So we add a 2-clause $C_{ij}$ that requires $x_ix_j$ to take one of these
values and take $T = Q\ann_{i,j} C_{ij}$.
Clearly, every assignment along the path lies in $T$.
We claim that these are the only solutions. To
show this, take an arbitrary assignment $\mb{c}$ satisfying the added
constraints. If for some $i < j \leq r$ we have $c_i = a_i$ but $c_j =
\bar{a}_j$, this would violate $C_{ij}$.
Hence the first $r$ variables of $\mb{c}$ are of the form
$\bar{a}_1\dots\bar{a}_ia_{i+1}\dots a_r$ for $0 \leq i
\leq r$. If $c_{r+1} =\bar{a}_{r+1}$ then $\mb{c} = \mb{u}^i$. If
$c_{r+1} = a_{r+1}$ then $\mb{c} = \mb{u}^i \oplus \mb{e}_{r+1}$. By
property 4 above, such a vector satisfies $Q$ if and only if $i =0$ or $i =r$,
which correspond to $\mb{c} = \mb{a}$ and $\mb{c} = \mb{b}$
respectively.
%\end{proof}

\begin{step}
\label{step:3}
Faithfully expressing paths of length $4$.
\end{step}
Let $\cal{P}$ denote the set of all ternary relations whose graph is a
path of length $4$ between two assignments at Hamming distance $2$. Up
to permutations of coordinates, there are 6 such relations. Each of
them is the conjunction of a $3$-clause and a $2$-clause. For
instance, the relation $M = \{100, 110, 010, 011, 001\}$ can be
written as $(x_1 \orr x_2 \orr x_3) \ann (\bar{x}_1 \orr
\bar{x}_3)$.  (It is named so, because its graph looks like the letter
'M' on the cube.) These relations are ``minimal" examples of relations
that are not componentwise bijunctive. By projecting out intermediate
variables from the path $T$ obtained in Step \ref{step:2}, we
faithfully express one of the relations in $\cal{P}$. We faithfully
express other relations in $\cal{P}$ using this relation.

We will write all relations in $\cal{P}$ in terms of $M(x_1,x_2,x_3) = (x_1
\orr x_2 \orr x_3) \ann (\bar{x}_1 \orr \bar{x}_3)$, by negating variables.
For example $M(\bar{x}_1, x_2,
x_3)=(\bar{x}_1 \orr x_2 \orr x_3) \ann (x_1 \orr \bar{x}_3) = \{000,
010, 110, 111, 101\}$.

\comment{
\begin{lemma}
\label{lem:r-to-2} Every relation  in $P$ is faithfully expressible from
$S$.
\end{lemma}
\begin{proof}
}

%Consider the path $T$ as above. If $r=2$, then $T\in P$. Suppose $r>2$.
Define the relation $P(x_1, x_{r+1}, x_2) = \exists x_3 \dots
x_r~T(x_1, \dots,x_{r+1})$.  The table below listing all tuples in $P$
and their witnesses, shows that the conditions for faithful
expressibility are satisfied, and $P \in \cal{P}$.

\vspace{0.6cm}
\begin{tabular}{|c|l|}
\hline
$x_1,x_2,x_{r+1}$ & $x_3,\dots, x_r$\\
\hline
$a_1a_2a_{r+1}$ & $a_3\dots a_r$\\
$a_1a_2\bar{a}_{r+1}$ &$a_3\dots a_r$\\
$\bar{a}_1a_2\bar{a}_{r+1}$ & $a_3\dots a_r$\\
$\bar{a}_1\bar{a}_2\bar{a}_{r+1}$ & $a_3\dots
a_k, \ \bar{a}_3a_4\dots a_r, \
\bar{a}_3\bar{a}_4a_5\dots a_r \ \dots
\bar{a}_3\bar{a}_4 \dots \bar{a}_r $\\
$\bar{a}_1\bar{a}_2a_{r+1}$ &
$\bar{a}_3 \bar{a}_4\dots \bar{a}_r$\\
\hline
\end{tabular}
\vspace{0.6cm}

Let $P(x_1, x_2,x_3) = M(l_1, l_2, l_3)$, where $l_i$ is one of $\{x_i,
\bar{x}_i\}$. We can now use $P$ and 2-clauses to express every
other relation in $\cal{P}$. Given $M(l_1, l_2, l_3)$ every relation
in $\cal{P}$ can be obtained by negating some subset of the variables.
Hence it suffices to show that we can express faithfully $M(\bar{l}_1,
l_2, l_3)$ and $M(l_1,
\bar{l}_2, l_3)$ ($M$ is symmetric in $x_1$ and $x_3$).
In the following let $\lambda$ denote one of the literals $\{y, \bar{y}\}$,
such that it is $\bar{y}$ if and only if $l_1$ is $\bar{x}_1$.
\begin{eqnarray*}
M(\bar{l}_1, l_2, l_3)
&= &  (\bar{l}_1 \orr l_2 \orr l_3) \ann (l_1 \orr \bar{l}_3)\\
& = & \exists y~(\bar{l}_1 \orr \bar{\lambda}) \ann
(\lambda \orr l_2 \orr l_3) \ann (l_1 \orr \bar{l}_3) \\
& = &  \exists y~(\bar{l}_1 \orr \bar{\lambda}) \ann
(\lambda \orr l_2 \orr l_3) \ann (l_1 \orr \bar{l}_3) \ann
(\bar{\lambda} \orr \bar{l}_3) \\
& = &  \exists y~(\bar{l}_1 \orr \bar{\lambda}) \ann
(l_1 \orr \bar{l}_3) \ann M(\lambda, l_2, l_3)\\
& = & \exists y~(\bar{l}_1 \orr \bar{\lambda}) \ann
(l_1 \orr \bar{l}_3) \ann P(y, x_2, x_3)
\end{eqnarray*}
In the second step the clause $(\bar{\lambda} \orr \bar{l}_3)$ is implied by 
the resolution of the clauses 
$(\bar{l}_1 \orr \bar{\lambda}) \ann (l_1 \orr \bar{l}_3)$.
 
For the next expression let $\lambda$ denote one of the literals $\{y,
\bar{y}\}$, such that it is negated if and only if $l_2$ is $\bar{x}_2$.
\begin{eqnarray*}
M(l_1, \bar{l}_2, l_3)
& = & (l_1 \orr \bar{l}_2 \orr l_3) \ann (\bar{l}_1 \orr \bar{l}_3) \\
& = & \exists y~(l_1 \orr l_3 \orr \lambda) \ann (\bar{\lambda} \orr \bar{l}_2)
\ann (\bar{l}_1 \orr \bar{l}_3) \\
& = & \exists y~(\bar{\lambda} \orr \bar{l}_2) \ann M(l_1, \lambda, l_3)\\
& =  & \exists y~(\bar{\lambda} \orr \bar{l}_2) \ann P(x_1, y, x_3)
\end{eqnarray*}
The above expressions are both based on resolution and it is easy to
check that they satisfy the properties of faithful expressibility.
%\end{proof}

\begin{step}
\label{step:4}
Faithfully expressing ${\cal S}_3$.
\end{step}
We faithfully express $(x_1\lor x_2\lor x_3)$ from $M$ using a
formula derived from a gadget in \cite{HD02}. This gadget
expresses $(x_1 \orr x_2 \orr x_3)$ in terms of ``Protected \OR'',
which corresponds to our relation $M$.

\comment{
\begin{lemma}
\label{p-to-3sat} Every 3-clause is faithfully expressible from $S$.
\end{lemma}
\begin{proof}
}

%The following formula was derived from a gadget in
%\cite{HD02} for expressing $(x_1 \orr x_2 \orr x_3)$ in terms of
%``Protected \OR'', which corresponds to our relation $M$
\begin{eqnarray}
\label{eq:demaine}
(x_1 \orr x_2 \orr x_3) & = & \exists y_1\dots y_5~(x_1 \orr \bar{y}_1)
\ann (x_2 \orr \bar{y}_2) \ann (x_3 \orr \bar{y}_3)
\ann (x_3 \orr \bar{y}_4) \nonumber \\
& & \ann M(y_1, y_5, y_3) \ann M(y_2, \bar{y}_5, y_4)
\end{eqnarray}
The table below listing the witnesses of each assignment for $(x_1,x_2, x_3)$, 
shows that the conditions for faithful expressibility are satisfied.

\vspace{0.6cm}
\noindent
% for 3SAT from M
\begin{scriptsize}
\begin{tabular}{|c|llll|}
\hline
$x_1,x_2,x_3$ & $y_1 \dots y_5$ &&&\\
\hline
$111$ & $00011$ ~$00111$ ~$00110$ ~$00100$ ~$01100$ ~$01101$ 
      & $01001$ ~$11001$ ~$11000$ & $10000$ & $10010$ ~$10011$ \\
$110$ &&$01001$ ~$11001$ ~$11000$ & $10000$ &         \\
$100$ &&                          & $10000$ &         \\
$101$ & $00011$ ~$00111$ ~$00110$ ~$00100$
      &                           & $10000$ & $10010$ ~$10011$ \\
$001$ & $00011$ ~$00111$ ~$00110$ ~$00100$ &&&\\
$011$ & $00011$ ~$00111$ ~$00110$ ~$00100$ ~$01100$ ~$01101$ & $01001$&&\\
$010$ &&$01001$&&\\
\hline
\end{tabular}
\end{scriptsize}
 \vspace{0.6cm}

From the relation $(x_1 \vee x_2 \vee x_3)$ we derive
the other 3-clauses by reverse resolution, for instance
$$(\bar{x}_1 \vee x_2 \vee x_3) = \exists y~(\bar{x}_1 \vee
\bar{y})\wedge (y \vee x_2 \vee x_3)$$
%\end{proof}

\end{proof}

To complete the proof of the \FET, we show that an arbitrary relation
can be expressed faithfully from ${\cal S}_3$.

\begin{lemma}
\label{lem:R} Let $R \subseteq \{0,1 \}^k$ be any relation of arity
$k\ge 1$. $R$ is faithfully expressible from ${\cal S}_3$.
\end{lemma}

\begin{proof}
If $k\le 3$ then $R$ can be expressed as a formula in \cnf$({\cal
S}_3)$ with constants, without introducing witness variables. This
kind of expression is always faithful.

If $k\ge 4$ then $R$ can be expressed as a formula in $\cnf({\cal
S}_k)$, without witnesses (i.e. faithfully). We will show that every
$k$-clause can be expressed faithfully from ${\cal S}_{k-1}$. Then, by
induction, it can be expressed faithfully from ${\cal S}_3$. For
simplicity we express a $k$-clause corresponding to the relation
$D_0$. The remaining relations are expressed equivalently.  We
express $D_0$ in a way that is standard in other complexity
reductions, and turns out to be faithful:
$$(x_1 \vee x_2 \vee \dots \vee x_k) =
\exists y ~
(x_1 \vee x_2 \vee y) \wedge (\bar{y} \vee x_3 \vee \dots \vee x_k).$$
This is the reverse operation of resolution.  For any satisfying
assignment for $\mb{x}$, its witness space is either $\{0\}$, $\{1\}$
or $\{0,1\}$, so in all cases it is connected. Furthermore, the only
way two neighboring satisfying assignments for $x$ can have no common
witness is if one of them has witness set $\{0\}$, and the other one
has witness set $\{1\}$. This implies that the first one has
$(x_3,\dots,x_k) = (0,\dots,0)$, and the other one has $(x_1,
x_2)=(0,0)$, thus they differ in the assignments of at least two
variables: one from $\{x_1, x_2\}$ and one from $\{x_3, \dots, x_k
\}$. In that case they cannot be neighboring assignments. Therefore all
requirements of faithful expressibility are satisfied.
\end{proof}

\subsection{Hardness Results for $3$-\cnf~formulas}
\label{sec:3sat}

%phk
From Lemma \ref{lem:expr} and Corollary \ref{cor:red}, it follows
that, to prove the hard side of our dichotomy theorems, it
suffices  to focus on $3$-\cnf~ formulas.

The proof that \conn$({\cal S}_3)$ and \stconn$({\cal S}_3)$ are
\PSPACE-complete is fairly intricate; it entails  a direct
%chp
reduction from the computation of a space-bounded Turing
machine. The result for \stconn~can also be proved easily using results
of Hearne and Demaine on Non-deterministic Constraint Logic
\cite{HD02}. However, it does not appear that completeness for
\conn~follows from their results.

\begin{lemma}
\label{lem:machine} \stconn$({\cal S}_3)$ and  \conn$({\cal S}_3)$
are \PSPACE-complete.
\end{lemma}
\begin{proof}
Given a \cnf(${\cal S}_3$) formula $\varphi$ and satisfying
assignments $\mb{s}, \mb{t}$ we can check if they are connected in $G(\varphi)$
with polynomial amount of space. Similarly for \conn(${\cal S}_3$), by
reusing space we can check for all pairs of assignments whether they
are satisfying and, if they both are, whether they are connected in
$G(\varphi)$.  It follows that both problems are in \PSPACE.

%chp modified the following by assuming no input
Next we show that \conn$({\cal S}_3)$ and \stconn$({\cal S}_3)$ are
\PSPACE-hard. Consider the following known PSPACE-complete problem:
Given a deterministic Turing machine $M=(Q,\Sigma, \Gamma, \delta,
q_0, \qaccept, \qreject)$ and $n$ in unary, will $M$ accept the string
consisting of $n$ blanks, without ever leaving its $n$ tape squares?
We give a polynomial time reduction from this problem to 
\stconn$({\cal S}_3)$ and \conn$({\cal S}_3)$.

The reduction maps a machine $M$ and integer $n$ (without loss of generality, assuming that
$n$ is at least as large as the description of $M$) to a $3$-\cnf~formula
$\varphi$ and two satisfying assignments for the formula, which are
connected in $G(\varphi)$ if and only if $M$ accepts. Furthermore,
all satisfying assignments of $\varphi$ are connected to one of these two
assignments, so that $G(\varphi)$ is connected if and only if $M$ accepts
$w$.

Before we show how to construct $\varphi$, we modify $M$ in several
ways:
\begin{enumerate}
\item We add a clock that counts from $0$ to $n\times |Q|\times
|\Gamma|^{n} = 2^{O(n)}$, which is the total number of
possible distinct configurations of $M$. It uses a separate tape of
length $O(n)$ with the alphabet $\{0,1\}$. Before a
transition happens, control is passed on to the clock, its counter is
incremented, and finally the transition is completed.

\item We define a standard accepting configuration.  Whenever $\qaccept$ is
reached, the clock is stopped and set to zero, the original tape
is erased and the head is placed in the initial position, always in state
$\qaccept$.

\item Whenever $\qreject$ is reached the machine goes into its initial
configuration. The tape is erased, the clock is set to zero, the head is
placed in the initial position, and the state is set to $q_0$ (and thus the
computation resumes).

\item Whenever the clock overflows, the machine goes into $\qreject$.
\end{enumerate}

The new machine $M'$ runs forever if $M$ does not accept (rejects or loops), 
and accepts if $M$ accepts. It also has the property that every configuration leads
either to the accepting configuration or to the initial configuration.  
Therefore the space of configurations is connected if
and only if $M$ accepts. Let's denote by $Q'$ the states of $M'$ and by
$\delta'$ its transitions. $M'$ runs on two tapes, the main one of size $N$
and the clock of size $N_c$, both $O(n)$. 
The alphabet of $M'$ on one tape is $\Gamma$, and on
the other $\{0,1\}$. For simplicity we can also assume
that at each transition the machine uses only one of the two tapes.

Next, we construct an intermediate \cnf-formula $\psi$ whose solutions are the
configurations of $M'$. However, the space of solutions of $\psi$ is
disconnected.

For each $i \in [N]$ and $a \in \Gamma$, we have a variable $x(i,
a)$. If $x(i, a) =1$, this means that the $i^{th}$ tape cell contains
symbol $a$. For every $i\in [N]$ there is a variable $y(i)$ which is 1
if the head is at position $i$. For every $q \in Q'$, there is a
variable $z(q)$ which is 1 if the current state is $q$.  Similarly
for every $j \in [N_c]$ and $a \in \{0,1\}$ we have variables $x_c(j,
a)$ and a variable $y_c(j)$ which is 1 if the head of the clock tape
is at position $j$.

We enforce the following conditions:
\begin{enumerate}
\item Every cell contains some symbol:
$$ \psi_1 = \bigwedge_{i \in [N]} \left(\vee_{a \in \Gamma}~~x(i,a)\right)
\bigwedge_{j \in [N_c]} \left(\vee_{a \in \{0,1\}}~~x_c(j,a)\right).$$
\item No cell contains two symbols:
$$ \psi_2 = \bigwedge_{i \in [N]} \bigwedge_{a \neq a' \in \Gamma}
\left(\overline{x(i, a)} \vee \overline{x(i,a')}\right)
\bigwedge_{j \in [N_c]}
\left(\overline{x_c(j, 0)} \vee \overline{x_c(j,1)}\right).$$
\item The head is in some position, the clock head is in some position, and the machine is in some state:
$$\psi_3 = \left(\vee_{i\in [N]}~~y(i)\right) \bigwedge
\left(\vee_{j\in [N_c]}~~y_c(j)\right) \bigwedge
\left(\vee_{q \in Q_1}~~z(q)\right).$$
\item The main tape head is in a unique position, the clock head is in a unique position, and the machine is in a unique state:
$$ \psi_4=
\bigwedge_{i \neq i' \in [N]} \left(\overline{y(i)} \vee \overline{y(i')} \right)
\bigwedge_{j \neq j' \in [N_c]} \left(\overline{y_c(j)} \vee \overline{y_c(j')} \right)
\bigwedge_{q \neq q' \in Q'} \left( \overline{z(q)} \vee \overline{z(q')} \right)
.$$
\end{enumerate}

Solutions of $\psi = \psi_1 \ann \psi_2 \ann \psi_3 \ann
\psi_4$ are in 1-1 correspondence with configurations of $M'$. Furthermore, the
assignments corresponding to any two distinct configurations differ
in at least two variables (hence the space of solutions is totally disconnected).

Next, to connect the solution space along valid transitions of $M'$,
we relax conditions 2 and 4 by introducing new transition variables,
which allow the head to have two states or a cell to have two symbols
at the same time. This allows us to go from one configuration to the next.

Consider a transition $\delta(q, a) = (q', b, R)$, which operates on
the first tape, for example. Fix the position of the head of the first
tape to be $i$, and the symbol in position $i+1$ to be $c$. The
variables that are changed by the transition are: $x(i, a)$, $y(i)$, $z(q)$,
$x(i, b)$, $y(i+1)$, $z(q')$. Before the transition the first three are set
to 1, the second three are set to 0, and after the transition they are
all flipped.  Corresponding to this transition (which is specified by
$i$, $q$, $a$, and $c$) we introduce a transition variable
$t(i,q,a,c)$. We now relax conditions 2 and 4 as follows:

\begin{itemize}
\item Replace $\left(\overline{x(i, a)} \vee \overline{x(i,b)}\right)$ by
$\left(\overline{x(i, a)} \vee \overline{x(i,b)} \vee t(i,q,a,c)\right)$.
\item Replace $\left(\overline{y(i)} \vee \overline{y(i+1)}\right)$ by
$\left(\overline{y(i)} \vee \overline{y(i+1)} \vee t(i,q,a,c)\right)$.
\item Replace $\left(\overline{z(q)} \vee \overline{z(q')}\right)$ by
$\left(\overline{z(q)} \vee \overline{z(q')} \vee t(i,q,a,c)\right)$.
\end{itemize}

This is done for every value of $q$, $a$, $i$ and $c$ (and also for
transitions acting on the clock tape). We add the transition variables
to the corresponding clauses so that for example the clause
$\left(\overline{x(i,a)} \vee \overline{x(i,b)}\right)$ could
potentially become very long, such as: 
$$\left(\overline{x(i,a)} \vee
\overline{x(i,b)} \vee t(i,q_1, a, c_1) \vee t(i, q_2, a, c_2) \vee
\dots \right).$$ 
However, the total number of transition variables is only
polynomial in $n$. We also add a constraint for every pair of
transition variables $t(i,q,a,c)$, $t(i',q',a',c')$, saying they
cannot be 1 simultaneously: $(\overline{t(i,q,a,c)} \orr
\overline{t(i',q',a',c')})$.  This ensures that only one transition
can be happening at any time.  The effect of adding the transition
variables to the clauses of $\psi_2$ and $\psi_4$ is that by setting
$t(i,q,a,c)$ to 1, we can simultaneously set $x(i,a)$ and $x(i,b)$ to
$1$, and so on.  This gives a path from the initial configuration to
the final configuration as follows: Set $t(i,q,a,c) =1$, set $x(i,b) =1$,
$y(i+1)=1$, $z(q')=1$, $x(i,a) =0$, $y(i)=0$, $z(q)=0$, then set
$t(i,q,a,c)=0$. Thus consecutive configurations are now connected. To avoid
connecting to other configurations, we also add an expression to
ensure that these are the only assignments the 6 variables can take
when $t(i,q,a,c)=1$:
\begin{eqnarray*}
\psi_{i,q,a,c} &=& \overline{t(i,q,a,c)} \orr
((x(i,a),y(i),z(q), x(i,b)), y(i+1), z(q')) \in \\
&&~~~~~~~~~~~~~~~~~~~~~~~~\{111000, 111100, 111110,
111111, 011111, 001111, 000111\}).
\end{eqnarray*}
This expression can of course be written in conjunctive normal form.

Call the resulting \cnf~formula $\varphi(\bf{x,x_c, y, y_c,
z,t})$. Note that $\varphi({\bf x, x_c, y, y_c, z, 0}) = \psi({\bf x,
x_c,y, y_c, z})$, so a solution where all transition variables are $0$
corresponds to a configuration of $M'$. To see that we have not
introduced any shortcut between configurations that are not valid
machine transitions, notice that in any solution of $\varphi$, at most
a single transition variable can be $1$. Therefore none of the
transitional solutions belonging to different transitions can be
adjacent. Furthermore, out of the solutions that have a transition
variable set to 1, only the first and the last correspond to a valid
configuration. Therefore none of the intermediate solutions can be
adjacent to a solution with all transition variables set to 0.

The formula $\varphi$ is a \cnf~formula where clause size is
unbounded. We use the same reduction as in the proof of Lemma \ref{lem:R}
to get a 3-\cnf~formula.
By Lemma \ref{lem:red} and Corollary \ref{cor:red},
\stconn~ and \conn~ for ${\cal S}_3$ are
\PSPACE-complete.
\end{proof}

By Lemma \ref{lem:expr} and Corollary \ref{cor:red}, this completes
the proof of the hardness part of the dichotomies for \conn~ and \stconn~ (Theorems
\ref{thm:conn} and \ref{thm:stconn}).

%phk moved diameter to the end of this section

Finally, we show that $3$-\cnf~formulas can have exponential
diameter, by inductively constructing a path of length at least
$2^\frac{n}{2}$ on $n$ variables and then identifying it with the
solution space of a $3$-\cnf~formula with $O(n^2)$ clauses. By
Lemma \ref{lem:expr} and Corollary \ref{cor:red}, this implies the
hardness part of the diameter dichotomy (Theorem \ref{thm:diam}).

\begin{lemma}
\label{lem:longpath} For $n$ even, there is a $3$-\cnf~formula
$\varphi_n$ with $n$ variables and $O(n^2)$ clauses, such that
$G(\varphi_n)$ is a path of length greater than $2^\frac{n}{2}$.
\end{lemma}
\begin{proof}
The construction is in two steps: we first exhibit an induced
subgraph $G_n$ of the $n$ dimensional hypercube with large
diameter. We then construct a 3-\cnf~ formula $\varphi_n$ so that
$G_n = G(\varphi_n)$.

The graph $G_n$ is a path of length $2^\frac{n}{2}$. We construct
it using induction. For $n=2$, we take $V(G_2) = \{(0,0), (0,1),
(1,1)\}$ which has diameter $2$. Assume that we have constructed
$G_{n-2}$ with $2^\frac{n-2}{2}$ vertices, and with distinguished
vertices $\mb{s_{n-2}},\mb{t_{n-2}}$ such that the shortest path
from $\mb{s}$ to $\mb{t}$ in $G_{n-2}$ has length
$2^\frac{n-2}{2}$. We now describe the set $V(G_n)$. For each
vertex $\mb{v} \in V(G_{n-2})$, $V(G_n)$ contains two vertices
$(\mb{v}, 0,0)$ and $(\mb{v}, 1,1)$. Note that the subgraph
induced by these vertices alone consists of two disconnected
copies of $G_{n-2}$. To connect these two components, we add the
vertex $\mb{m} = (\mb{t},0,1)$ (which is connected to $(\mb{t},
0,0)$ and $(\mb{t},1,1)$ in the induced subgraph).  Note that the
resulting graph $G_n$ is connected, but any path from $(\mb{u},
0,0)$ to $(\mb{v}, 1, 1)$ must pass through $\mb{m}$. Further note
that by induction, the graph $G_n$ is also a path. The vertices
$\mb{s_n} = (\mb{s_{n-2}}, 0,0)$ and $\mb{t_n} =
(\mb{s_{n-2}},1,1)$ are diametrically opposite ends of this path.
The path length is at least $2\cdot 2^\frac{n-2}{2} +2 >
2^\frac{n}{2}$. Also $\mb{s_2} = (0,0), \ \ \mb{s_{n}} =
(\mb{s_{n-2}}, 0, 0), \ \ \mb{t_n} = (\mb{s_{n-2}},1,1)$ and hence
$\mb{s_n} = (0,\dots, 0), \mb{t_n} = (0, \dots, 0,1,1)$.

We construct a sequence of 3-\cnf~formulas $\varphi_n(x_1, \dots,
x_n)$ so that $G_n = G(\varphi_n)$. Let $\varphi_2(x_1, x_2) =
\bar{x}_1 \vee x_2$. Assume we have $\varphi_{n-2}(x_1, \dots,
x_{n-2})$. We add two variables $x_{n-1}$ and $x_n$ and the
clauses
\begin{eqnarray}
&\varphi_{n-2}(x_1, \dots, x_{n-2}), \ \ \bar{x}_{n-1} \wedge x_n & \nonumber \\
&x_{n-1} \vee \bar{x}_n \vee \bar{x}_i & \text{for} \ \ i \leq n-4 \label{eqn:3} \\
&x_{n-1} \vee \bar{x}_n \vee x_i &  \text{for} \ \ i = n-3, n-2
\label{eqn:4}
\end{eqnarray}
Note that a clause in \ref{eqn:3} is just the implication
$(\bar{x}_{n-1} \wedge x_n) \rightarrow \bar{x}_i$. Thus clauses
\ref{eqn:3}, \ref{eqn:4} enforce the condition that $x_{n-1} = 0,
x_n =1$ implies that $(x_1, \dots, x_{n-2}) = \mb{t_{n-2}} = (0,
\dots, 0, 1, 1)$.
\end{proof}

\section{The Easy Case of the Dichotomy: Tight Sets of Relations}
\label{sec:easy}

\subsection{Schaefer sets of relations}

We begin by showing that all Schaefer sets of relations are
tight. Schaefer relations are characterized by closure
properties. We say that a $r$-ary relation $R$ is closed
under some $k$-ary operation $\alpha:\{0,1\}^k \rightarrow \{0,1\}$ if
for every $\mb{a^1},
\mb{a^2}, \dots, \mb{a^k} \in R$, the tuple $(\alpha(a^1_1, a^2_1,
\dots, a^k_1), \dots, \alpha(a^1_r, \dots, a^k_r))$ is in $R$. We
denote this tuple by $\alpha(\mb{a^1}, \dots, \mb{a^k})$.

We will use the following lemma about closure properties on several
occassions.

\begin{lemma} \label{lem:idemp}
If a logical relation $R$ is closed under an operation
$\alpha: \{0,1\}^k \rightarrow \{0,1\}$ such that
$\alpha(1,1,\dots, 1)=1$ and $\alpha(0,0, \dots, 0)=0$ (a.k.a. an idempotent
operation) then every connected component of $G(R)$ is closed under $\alpha$.
\end{lemma}
\begin{proof}
Consider $\mb{a^1}, \dots, \mb{a^k} \in R$, such that they all belong
to the same connected component of $G(R)$. It suffices to prove that
$\mb{a}=\alpha(\mb{a^1}, \dots, \mb{a^k})$ is in the same connected component
of $G(R)$. To that end we will first prove that for any $\mb{s},
\mb{t} \in R$ if there is a path
from $\mb{s}$ to $\mb{t}$ in $G(R)$ then there is a path from
$\alpha(\mb{b^1}, \dots, \mb{b^{i-1}},\mb{s}, \mb{b^{i+1}},
\dots, \mb{b^k})$
to
$\alpha(\mb{b^1}, \dots, \mb{b^{i-1}},\mb{t}, \mb{b^{i+1}},
\dots, \mb{b^k})$
for any $\mb{b^1}, \dots, \mb{b^k} \in R$. This observation implies
that there is a path from
$\mb{a^1}=\alpha(\mb{a^1}, \mb{a^1}, \dots, \mb{a^1})$ to
$\alpha(\mb{a^1},\mb{a^2}, \mb{a^1}, \dots, \mb{a^1})$, from there to
$\alpha(\mb{a^1},\mb{a^2}, \mb{a^3}, \mb{a^1}, \dots, \mb{a^1})$ and
so on, to
$\alpha(\mb{a^1},\mb{a^2}, \dots, \mb{a^k})=\mb{a}$. Thus $\mb{a}$ is
in the same connected component of $G(R)$ as $\mb{a^1}$.

Let the path from $\mb{s}$ to $\mb{t}$ be
$\mb{s}=\mb{s^1} \rightarrow \mb{s^2} \rightarrow \dots \mb{s^m}=\mb{t}$.
For every $j\in \{1, 2, \dots, m-1\}$, the tuples
$\alpha(\mb{b^1}, \dots, \mb{b^{i-1}},\mb{s^j},
\mb{b^{i+1}}, \dots, \mb{b^m})$
and
$\alpha(\mb{b^1}, \dots, \mb{b^{i-1}},\mb{s^{j+1}},
\mb{b^{i+1}}, \dots, \mb{b^m})$ differ in at most one position (the
position in which $\mb{s^j}$ and $\mb{s^{j+1}}$ are different)
therefore they belong to the same component of $G(R)$.
Thus $\alpha(\mb{b^1}, \dots, \mb{b^{i-1}},\mb{s^1},
\mb{b^{i+1}}, \dots, \mb{b^m})$ and
$\alpha(\mb{b^1}, \dots, \mb{b^{i-1}},\mb{s^m},
\mb{b^{i+1}}, \dots, \mb{b^m})$ belong to the same component.
\end{proof}

We are ready to prove that all Schaefer relations are tight.

\begin{lemma} \label{lem:relationship} Let $R$ be a logical relation.
\begin{enumerate}
\item If $R$ is bijunctive, then $R$ is componentwise
bijunctive.
\item If $R$ is Horn,  then $R$ is  \OR-free.
\item If $R$ is dual Horn, then $R$ is  \NAND-free.
\item If R is affine, then $R$ is componentwise
bijunctive, \OR-free, and \NAND-free.
\end{enumerate}
\end{lemma}

\begin{proof}
The case of bijunctive relations follows immediately from Lemma
\ref{lem:idemp} and the fact that a relation is bijunctive if and only if
it is closed under the ternary majority operation $\maj$, which is idempotent.

The cases of Horn and dual Horn are symmetric. Suppose a r-ary Horn relation
$R$ is not \OR-free. Then there exist $i,j \in \{1,\dots, r\}$ and
constants $t_1, \dots,t_r \in \{0,1\}$ such that the 
relation 
$R(t_1, \dots, t_{i-1}, x, t_{i+1}, \dots, t_{j-1}, y, t_{j+1},
\dots, t_r)$
on variables $x$ and $y$ is equivalent to $x\vee y$, i.e.
$$R(t_1, \dots, t_{i-1}, x, t_{i+1}, \dots, t_{j-1}, y, t_{j+1},
\dots, t_r) = \{01,11,10\}.$$
Thus the tuples $\mb{t^{00}}, \mb{t^{01}}\mb{t^{10}}, \mb{t^{11}}$
defined by
$(t_i^{ab}, t_j^{ab})=(a,b)$  and
$t_k^{ab}= t_k$ for every $k\not\in\{i,j\}$, where $a,b, \in \{0,1\}$
satisfy $\mb{t^{10}}, \mb{t^{11}}, \mb{t^{01}} \in R$
and $\mb{t^{00}}\not\in R$.
However, since every Horn relation is closed under $\wedge$, it follows that
$\mb{t^{01}} \wedge \mb{t^{10}}=\mb{t^{00}}$ must be in $R$, which is
a contradiction.

For the affine case, a small modification of the last step of the
above argument shows that an affine relation also is \OR-free;
therefore, dually, it is also
\NAND-free. Namely, since a relation $R$ is affine if and only if it is
closed under
ternary $\oplus$, it follows that $\mb{t^{01}} \oplus \mb{t^{11}} \oplus
\mb{t^{10}}=\mb{t^{00}}$ must be in $R$.

Since the connected components of an affine relation are both \OR-free
and \NAND-free the subgraphs that they induce are hypercubes, which
are also bijunctive relations. Therefore an affine relation is also
componentwise bijunctive.
\end{proof}

%It is not hard to show that
These containments are proper. For instance, $R_{1/3}= \{100, 010,
001\}$ is componentwise bijunctive, but  not bijunctive as
$\maj(100,010, 001) = 000 \not \in R_{1/3}$.

\subsection{Structural properties of tight sets of relations}

In this section, we explore some structural properties of the
solution graphs of tight sets of relations. These properties provide
simple algorithms for \conn$({\cal S})$ and \stconn$({\cal S})$ for tight
sets ${\cal S}$, and also guarantee that for such sets, the diameter of
$G(\varphi)$ of \cnf$({\cal S})$-formula $\varphi$ is linear.

\begin{lemma}
\label{lem:2sat} Let ${\cal S}$ be a set of componentwise bijunctive
relations and $\varphi$ a \cnf$({\cal S})$-formula. If $\mb{a}$ and
$\mb{b}$ are two solutions of $\varphi$ that lie in the same
component of $G(\varphi)$, then
 $d_\varphi(\mb{a}, \mb{b}) = |\mb{a} -\mb{b}|$.
\end{lemma}
\begin{proof}
Consider first the special case in which every relation in ${\cal S}$ is
bijunctive. In this case, $\varphi$ is equivalent to a 2-\cnf~formula
and so the space of solutions of $\varphi$ is closed under majority.  We
show that there is a path in $G(\varphi)$ from $\mb{a}$ to $\mb{b}$,
such that along the path only the assignments on variables with
indices from the set $D=\{i| a_i \neq b_i\}$ change. This implies that
the shortest path is of length $|D|$ by induction on $|D|$. Consider
any path $\mb{a}
\rightarrow \mb{u^1}
\rightarrow \dots
\rightarrow \mb{u^r} \rightarrow \mb{b}$ in $G(\varphi)$.
We construct another path by replacing $\mb{u^i}$ by
$\mb{v^i}=\maj~(\mb{a}, \mb{u^i}, \mb{b})$ for $i=1, \dots, r$, and
removing repetitions. This is a path because for any $i$ $\mb{v^i}$
and $\mb{v^{i+1}}$ differ in at most one variable. Furthermore,
$\mb{v^{i}}$ agrees with $\mb{a}$ and $\mb{b}$ for every $i$ for which
$a_i=b_i$. Therefore, along this path only variables in $D$ are flipped.

For the general case, we show that every  component $F$
of $G(\varphi)$ is the solution space of a 2-\cnf~formula
$\varphi'$. Let $F$ be the component of $G(\varphi)$ which
contains $\mb{a}$ and $\mb{b}$. Let $R \in {\cal S}$ be a relation with
two components, $R_1, R_2$ each of which are bijunctive.
Consider a clause in $\varphi$ of the form $R(x_1, \dots, x_k)$.
The projection of $F$ onto $x_1, \dots, x_k$ is itself
connected and must satisfy $R$. Hence it lies within one of the
two components $R_1, R_2$, assume it is $R_1$. We replace $R(x_1,
\dots, x_k)$ by $R_1(x_1, \dots, x_k)$. Call
this new formula $\varphi_1$. $G(\varphi_1)$ consists of all
components of $G(\varphi)$ whose projection on $x_1,
\dots, x_k$ lies in $R_1$. We repeat this for every clause. Finally
we are left with a formula $\varphi'$ over a set of bijunctive
relations. Hence $\varphi'$ is bijunctive and $G(\varphi')$ is a
component of $G(\varphi)$. So the claim follows from the bijunctive case.
\end{proof}

\begin{corollary}
\label{cor:2sat} Let ${\cal S}$ be a set of componentwise bijunctive
relations. Then
\begin{enumerate}
\item For every $\varphi \in \cnf({\cal S})$ with $n$ variables, the
diameter of each  component of $G(\varphi)$ is bounded by $n$.
\item \stconn$({\cal S})$ is in \PTIME.
\item \conn$({\cal S})$ is in \coNP.
\end{enumerate}
\end{corollary}
\begin{proof}
The bound on diameter is an immediate consequence of Lemma
\ref{lem:2sat}.

The following algorithm solves \stconn$({\cal S})$ given
vertices $\mb{s}, \mb{t} \in G(\varphi)$. Start with $\mb{u} =
\mb{s}$. At each step, find a variable $x_i$ so that $u_i \neq t_i$
and flip it, until we reach $\mb{t}$. If at any stage no such variable
exists, then declare that $\mb{s}$ and $\mb{t}$ are not connected.  If
the $\mb{s}$ and $\mb{t}$ are disconnected, the algorithm is bound to
fail.  So assume that they are connected. Correctness is proved by
induction on $d = |\mb{s} - \mb{t}|$.  It is clear that the algorithm
works when $d =1$. Assume that the algorithm works for $d-1$. If $s$
and $t$ are connected and are distance $d$ apart, Lemma \ref{lem:2sat}
implies there is a path of length $d$ between them in $G(\varphi)$. In
particular, the algorithm will find a variable $x_i$ to flip. The
resulting assignment is at distance $d-1$ from $\mb{t}$, so now we
proceed by induction.

Next we prove that \conn$({\cal S})$ $\in$ \coNP. A short certificate
that the graph is not connected is a pair of assignments $\mb{s}$ and $\mb{t}$ which
are solutions from different components. To verify that they are
disconnected it suffices to run the algorithm for \stconn.
\end{proof}

We consider sets of \OR-free relations.
Define the \emph{coordinate-wise partial order} $\leq$ on Boolean
vectors as follows: $\mb{a} \leq
\mb{b}$ if $a_i \leq b_i$, for each  $i$.

\begin{lemma}
\label{lem:or-free} Let ${\cal S}$ be a set of \OR-free relations and
$\varphi$  a \cnf$({\cal S})$-formula. Every component of
$G(\varphi)$ contains a minimum solution with respect to the
coordinate-wise order; moreover, every solution is connected to the
minimum solution in the same component via a monotone path.
\end{lemma}
\begin{proof}
We call a satisfying assignment locally minimal, if it has no
neighboring satisfying assignments that are smaller than it. We will
show that there is exactly one such assignment in each component of
$G(\varphi)$.

Suppose there are two distinct locally minimal
assignments $\mb{u}$ and $\mb{u'}$ in some component of
$G(\varphi)$. Consider the path between them where the maximum Hamming
weight of assignments on the path is minimized. If there are many such
paths, pick one where the smallest number of assignments have the
maximum Hamming weight. Denote this path by $\mb{u}=\mb{u^1}
\rightarrow \mb{u^2} \rightarrow
\dots
\rightarrow \mb{u^r}= \mb{u'}$. Let $\mb{u^i}$ be an assignment of
largest Hamming weight in the path. Then $\mb{u^i}\not = \mb{u}$ and
$\mb{u^i}\not = \mb{u'}$, since $\mb{u}$ and $\mb{u'}$ are locally
minimal. The assignments $\mb{u^{i-1}}$ and $\mb{u^{i+1}}$ differ in
exactly 2 variables, say, in $x_1$ and $x_2$. So
$\{u^{i-1}_1u^{i-1}_2,~ u^i_1u^i_2,~ u^{i+1}_1u^{i+1}_2\}= \{01,
11,10\}$.  Let $\mb{\hat{u}}$ be such that $\hat{u}_1=\hat{u}_2=0$,
and $\hat{u}_i=u_i$ for $i>2$. If $\mb{\hat{u}}$ is a solution, then
the path $\mb{u^1}\rightarrow \mb{u^2} \rightarrow
\dots \rightarrow \mb{u^i} \rightarrow \mb{\hat{u}} \rightarrow \mb{u^{i+1}}
\rightarrow \dots
\rightarrow \mb{u^r}$
contradicts the way we chose the original path. Therefore,
$\mb{\hat{u}}$ is not a solution. This means that there is a
clause that is violated by it, but is satisfied by
$\mb{u^{i-1}}$, $\mb{u^i}$, and $\mb{u^{i+1}}$. So the relation
corresponding to that clause is not \OR-free, which is
a contradiction.

The unique locally minimal solution in a component is its minimum
solution, because starting from any other assignment in the component,
it is possible to keep moving to neighbors that are smaller, and the only time
it becomes impossible to find such a neighbor is when the locally minimal
solution is reached. Therefore, there is a monotone path from any
satisfying assignment to the minimum in that component.
\end{proof}

\begin{corollary}
\label{cor:or-free}
Let ${\cal S}$ be a set of \OR-free relations.  Then
\begin{enumerate}
\item For every $\varphi
\in \cnf({\cal S})$ with $n$ variables, the diameter of each component of
$G(\varphi)$ is bounded by $2n$.
\item \stconn$({\cal S})$ is in \PTIME.
\item \conn$({\cal S})$ is in \coNP.
\end{enumerate}
\end{corollary}
\begin{proof}
Given solutions $\mb{s}$ and $\mb{t}$ in the same component of
$G(\varphi)$, there is a monotone path from each to the minimal
solution $\mb{u}$ in the component. This gives a path from $\mb{s}$ to
$\mb{t}$ of length at most $2n$. To check if $\mb{s}$ and $\mb{t}$ are
connected, we just check that the minimal assignments reached from
$\mb{s}$ and $\mb{t}$ are the same.
\end{proof}

Sets of \NAND-free relations are handled dually to \OR-free
relations. In this case there is a maximum solution in every connected
component of $G(\phi)$ and every solution is connected to it via a
monotone path. Finally, putting everything together, we complete the proofs
of all our dichotomy theorems.

\begin{corollary}
\label{cor:tight}
Let ${\cal S}$ be a  tight set of relations.  Then
\begin{enumerate}
\item For every $\varphi
\in \cnf({\cal S})$ with $n$ variables, the diameter of each component of
$G(\varphi)$ is bounded by $2n$.
\item \stconn$({\cal S})$ is in \PTIME.
\item \conn$({\cal S})$ is in \coNP.
\end{enumerate}
\end{corollary}

\subsection{The Complexity of \conn~for Tight Sets of Relations}

We pinpoint the complexity of \conn$({\cal S})$ for the
tight cases which are not Schaefer, using a result of Juban
\cite{Juban99}.

\begin{lemma}
\label{lem:easy-CONN}
For ${\cal S}$ tight, but not Schaefer, \conn$({\cal S})$ is \coNP-complete.
\end{lemma}
\begin{proof}
The problem {\sc Another-Sat}$({\cal S})$ is: given a formula $\varphi$
in \cnf$({\cal S})$ and a solution $\mb{s}$, does there exist a
solution $\mb{t}\not = \mb{s}$? Juban (\cite{Juban99}, Theorem 2)
shows that if ${\cal S}$ is not Schaefer, then {\sc Another-Sat} is
\NP-complete.  He also shows (\cite{Juban99}, Corollary 1) that if ${\cal S}$ is not
Schaefer, then the relation $x \neq y$ is expressible from ${\cal S}$
through substitutions.

Since ${\cal S}$ is not Schaefer, {\sc Another-Sat}$({\cal S})$ is
\NP-complete. Let $\varphi, \mb{s}$ be an instance of {\sc
Another-Sat} on variables $x_1, \dots, x_n$. We define a
\cnf$({\cal S})$ formula $\psi$ on the variables $x_1, \dots, x_n,
y_1, \dots, y_n$ as $$\psi(x_1, \dots, x_n, y_1, \dots, y_n) =
\varphi(x_1, \dots, x_n) \wedge_i (x_i \neq y_i)$$ It is easy to
see that $G(\psi)$ is connected if and only if $\mb{s}$ is the
unique solution to $\varphi$.
\end{proof}

We are left with the task to determine the complexity of
\conn$({\cal S})$ for the case when ${\cal S}$ is a Schaefer set of relations.
In Lemmas \ref{lem:conn-2sat} and \ref{lem:conn-affine} we show that
\conn$({\cal S})$ is in \PTIME~ if ${\cal S}$ is affine or
bijunctive. This leaves the case of Horn and dual Horn, which we
discuss in the end of this section.

\begin{lemma} \label{lem:conn-2sat}
If ${\cal S}$ is a bijunctive set of relations then there is
a polynomial time algorithm for \conn$({\cal S})$.
\end{lemma}
\begin{proof}
Consider a formula $\phi(x_1, \dots, x_n)$ in \cnf$({\cal S})$. Since
${\cal S}$ is a bijunctive set of relations $\phi$ can be written as a
2-\cnf~formula. Since satisfiability of 2-\cnf~formulas is decidable
in polynomial time, it is easy to decide for a given variable $x_i$
whether there exist solutions in which it takes a particular value in
$\{0,1\}$.  The variables which can only take one value are assigned
that value. Without loss of generality we can assume that the
resulting 2-\cnf~formula is $\psi(x_1, \dots, x_m)$.

Consider the graph of implications of $\psi$ defined in the following
way: the vertices are the literals $x_1, \dots, x_m$, $\bar{x}_1, \dots,
\bar{x}_m$. There is a directed edge from literal $l_1$ to literal
$l_2$ if and only if $\psi$ contains a clause containing $l_2$ and the
negation of $l_1$, which we denote by $\bar{l}_1$ (if $l_1$ is a
negated variable $\bar{x}$, then $\bar{l}_1$ denotes $x$).  The
directed edge represents the fact that in a satisfying assignment if the
literal $l_1$ is assigned true, then the literal $l_2$ is also
assigned true. We will show that $G(\psi)$ is disconnected if and only
if the graph of implications contains a directed cycle. This property
can be checked in polynomial time.

Suppose the graph of implications contains a directed cycle of literals
$l_1\rightarrow l_2 \rightarrow l_3 \rightarrow \dots
\rightarrow l_k \rightarrow l_1$.
By the construction, the graph also contains a directed cycle on the negations
of these literals, but in the opposite direction:
$\bar{l}_k \rightarrow \bar{l}_{k-1} \rightarrow \dots \rightarrow
\bar{l}_2 \rightarrow \bar{l}_1 \rightarrow \bar{l}_k$.
There is a satisfying assignment $\mb{s}$ in which $l_1$ is assigned
1, and also a satisfying assignment $\mb{t}$ in which $\bar{l}_1$ is
assigned 1. By the implications, in $\mb{s}$ the literals $l_1, l_2,
\dots, l_k$ are assigned 1, and in \mb{t} $\bar{l}_1,
\bar{l}_2, \dots, \bar{l}_k$ are assigned 1. Suppose there is a path
from $\mb{s}$ to $\mb{t}$. Then let $l_i$ be the first literal in the
cycle whose value changes along the path from $\mb{s}$ to
$\mb{t}$. Then there is a satisfying assignment in which $l_i$ is
assigned $0$ whereas all other literals on the cycle are assigned
1. On the other hand, this cannot be a satisfying assignment because
the edge $(l_{i-1}, l_i)$ implies that there is a clause containing
only $l_i$ and the negation of $l_{i-1}$, and this clause is violated
by the assignment. This is a contradiction, therefore there can be no
path from $\mb{s}$ to $\mb{t}$.

Next, suppose the graph of implications contains no directed cycle,
and $G(\psi)$ is disconnected.
Let $\mb{s}$ and $\mb{t}$ be satisfying assignments from different
connected components of $G(\psi)$ that are at minimum Hamming
distance. Let $U$ be the set of variables on which $\mb{s}$ and
$\mb{t}$ differ. There are two literals corresponding to each
variable, and let $U^\mb{s}$ and $U^\mb{t}$ denote respectively the
literals that are true in $\mb{s}$ and in $\mb{t}$. The directed graph
induced by $U^\mb{s}$ in the implications graph contains no directed
cycle, therefore there exists a literal $l\in U^\mb{s}$ without an
incoming edge from a literal in $U^\mb{s}$. There is also no incoming
edge from any other true literal in $\mb{s}$, because $\mb{t}$ is also
satisfying. Thus the value of the
corresponding variable can be flipped and the resulting assignment is
still satisfying. This assignment is in the same component as $\mb{s}$
but it is closer to $\mb{t}$ which contradicts our choice of $\mb{s}$
and $\mb{t}$.
\end{proof}

\begin{lemma}\label{lem:conn-affine}
If ${\cal S}$ is an affine set of relations then there is
a polynomial time algorithm for \conn$({\cal S})$.
\end{lemma}
\begin{proof}
An affine formula can be described as the set of solutions of a linear
system of equations. For any solution, if only a variable that appears
in at least one of the equations is flipped, the resulting assignment
is not a solution. Therefore it suffices to check whether the system
has more than one solution (after variables that don't appear in any
equation are removed), which is easy by checking the rank of the
matrix obtained from the Gaussian elimination algorithm.
\end{proof}

We are left with characterizing the complexity of \conn~for sets of Horn
relations and for sets of dual Horn relations. 
%\pnote{Changed this a little} 
In the conference
version \cite{icalp} of the present paper, we had conjectured that if
${\cal S}$ is Horn or dual Horn, then \conn$({\cal S})$ is in \PTIME,
but this was disproved by Makino, Tamaki and
Yamamoto \cite{MTY07}. They showed that \conn$(\{R_2\})$ is
\coNP-complete, where $R_2=\{0,1\}^3\backslash\{110\}$, hence 
there exist Horn (and by symmetry also dual Horn) sets
of relations for which \conn~is \coNP-complete. 
Their proof is via a reduction from
{\sc Positive Not-All-Equal 3-\sat}, which as seen earlier is 
\sat$(\{R_{\rm {NAE}}\})$, where 
$R_{\rm{NAE}}=\{0,1\}^3\setminus \{000, 111\}$. 
This problem is also known as 3-Hypergraph 2-colorability, 

The relation $R_2$ is a 3-clause with one positive literal. We will
describe a natural set of Horn relations first introduced in \cite{CKS01}, 
which cannot be used to express $R_2$. We show that
for this set there is a polynomial time algorithm for \conn.

\begin{definition}
%{\rm 
A logical relation $R$ is \emph{implicative hitting
set-bounded$-$} or \emph{IHSB$-$} if it is the set of solutions of a Horn
formula in which all clauses of size greater than 2 have only negative
literals. Similarly, $R$ is \emph{implicative hitting
set-bounded$+$} or \emph{IHSB$+$} if it is the set of solutions of a
dual Horn formula in which all clauses of size greater than 2 have only
positive literals.
%}
\end{definition}

These types of logical relations can be characterized by closure
properties. A relation $R$ is IHSB$-$ if and only if it is closed under
$\mb{a} \wedge (\mb{b} \vee \mb{c})$; in other words if $\mb{a},
\mb{b}, \mb{c} \in R$, where $R$ is of arity $r$, then
$\mb{a} \wedge (\mb{b} \vee \mb{c})=
(a_1 \wedge (b_1 \vee c_1)~,~ a_2 \wedge (b_2 \vee c_2)~,~ \dots~,
a_r \wedge (b_r \vee c_r)) \in R$.
A relation $R$ is IHSB$+$ if and only if it is closed under
$\mb{a} \vee (\mb{b} \wedge \mb{c})$.
While the definition may at first look unnatural, it comes from Post's
classification of Boolean functions 
%\pnote{Need to explain, or add reference} 
(see \cite{BCRV04}).
One of the consequences of this
classification is that IHSB$-$ relations cannot express all Horn
relations, and in particular $R_2$, even in the sense of Schaefer's
expressibility. For the purposes of faithful expressibility we can
define an even larger class of relations which cannot faithfully
express $R_2$ (unless \PTIME \ =\ \coNP).

\begin{definition} %{\rm
A logical relation $R$ is \emph{componentwise IHSB$-$ (IHSB$+$)}
if every connected component of $G(R)$ is IHSB$-$ (IHSB$+$).
%}
\end{definition}

By Lemma \ref{lem:idemp}, every relation that is IHSB$-$ (IHSB$+$) is
also componentwise IHSB$-$ (IHSB$+$).  Of course, the class of
componentwise IHSB$-$ relations is much broader, and in fact includes
relations that are not even Horn, such as $R_{1/3}$, However in the
following lemma we are only considering componentwise IHSB$-$
(IHSB$+$) relations which are Horn (dual Horn).
%\pnote{Check, and add a nice example.}  
We will say that a set of relations ${\cal S}$ is componentwise
IHSB$-$ (IHSB$+$) if every relation in ${\cal S}$ is componentwise
IHSB$-$ (IHSB$+$).

\begin{lemma} If ${\cal S}$ is a set of relations that are Horn (dual Horn)
and componentwise IHSB$-$ (IHSB$+$), then there is a polynomial time
algorithm for \conn$({\cal S})$.
\end{lemma}
\begin{proof}
%\pnote{Have expanded the proof in many places}
First we consider the case in which every relation in ${\cal S}$ is
IHSB$-$. The formula can be written as a conjunction of Horn clauses,
such that clauses of length greater than 2 have only negative
literals.  Let all unit clauses be assigned and propagated---their
variables take the same value in all satisfying assignments. The
resulting formula is also IHSB$-$, and has two kinds of clauses:
2-clauses with one positive and one negative literal, and clauses of
size 2 or more with only negative literals. The assignment of zero to
all variables is satisfying. There is more than one connected
component if and only if there is another assignment that is locally
minimal by Lemma \ref{lem:or-free}. A locally minimal satisfying
assignment is such that if any of the variables assigned 1 is changed
to 0 the resulting assignment is not satisfying. Thus all variables
assigned 1 appear in at least one 2-clause with one positive and one
negative literal for which both variables are assigned 1. We say that
such an assignment certifies the disconnectivity.

To describe the algorithm, we first define the following implication
graph $G$. The vertices are the set of variables. There is a directed
edge $(x_i,x_j)$ if and only if $(x_j \vee \bar{x}_i)$ is a clause in
the IHSB$-$ representation. Let $S_1, \dots, S_m$ be the sets of
variables in clauses with only negative literals. 
For every variable $x_i$ let $T_i$ denote the set of variables
reachable from $x_i$ in the directed graph. Note that if $x_i$ is set
to $1$, then every variable in $T_i$ must also be set to $1$. 
The algorithm rejects if and only if there exists a variable $x_i$
such that $x_i \in T_i$ and $T_i$ does not contain $S_j$ for any $j
\in \{1, \dots, m\}$. We show that this happens if and only if the solution
graph is disconnected. Note that the algorithm runs in
polynomial time. 

Assume that the graph of solutions is disconnected and consider the
satisfying assignment $\mb{s}$ that certifies disconnectivity. Let $U$
be the set of variables $x_i$ such that $s_i =1$.  Since every
variable in $U$ appears in at least one 2-clause for which both
variables are from $U$, the directed graph induced by $U$ is such that
every vertex has an incoming edge. By starting at any vertex in $U$
and following the incoming edge backwards until we repeat some vertex,
we find a cycle in the subgraph induced by $U$. For any variable $x_i$
in such a cycle it holds that $x_i \in T_i$. Further $T_i \subseteq
U$, since setting $x_i$ to $1$ forces all variables in $T_i$ to be
$1$. Also $T_i$ cannot contain $S_j$ for any $j$, else the
corresponding clause would not be satisfied by $\mb{s}$. Thus the algorithm
rejects whenever the solution graph is disconnected.

Conversely, if the algorithm rejects, there exists a variable $x_i$
such that $x_i \in T_i$ and $T_i$ does not contain $S_j$ for any $j
\in \{1, \dots, m\}$. Consider the assignment in which all variables from
$T_i$ are assigned 1, and the rest are assigned 0. We will show that
this assignment is satisfying and it is a certificate for
disconnectivity.  Clauses which contain only negated variables are
satisfied since $S_j\not\subset T_i$ for all $j$. Now consider a
clause of the form $(x_j \vee \bar{x}_k)$ and note that there is a
directed edge $(x_k,x_j)$.  If $x_k =0$, this is satisfied. If $x_k =1$
then $x_k \in T_i$, and hence $x_j \in T_i$ because of the edge
$(x_k,x_j)$. But then $x_j$ is set to $1$, so the clause is
satisfied. To show that this solution is minimal, consider trying to
set $x_k \in T_i$ to $0$. There is an incoming edge $(x_j,x_k)$
for some $x_j \in T_i$, and hence a clause $(x_k \vee
\bar{x}_j)$, which will become unsatisfied if we set $x_k
=0$. Thus we have a certificate for the space being disconnected.

Next, consider a formula $\phi(x_1, \dots, x_n)$ in \cnf$({\cal S})$.
We reduce the connectivity question to one for a formula with IHSB$-$
relations.  Since satisfiability of Horn formulas is decidable in
polynomial time and every connected component of a Horn relation is a
Horn relation by Lemma \ref{lem:idemp},
it is easy to decide for a given clause and a given
connected component of its corresponding relation (the relation
obtained after identifying repeated variables), whether there exists a
solution for which the variables in this clause are assigned a value
in the specified connected component. If there exists a clause for
which there is more than one connected component for which solutions
exist, then the space of solutions is disconnected. This follows from
the fact that the projection of $G(\phi)$ on the hypercube
corresponding to the variables appearing in this clause is
disconnected.  Therefore we can assume that the relation corresponding
to every clause has a single connected component. Since that component
is IHSB$-$ the relation itself is IHSB$-$.
\end{proof}

It is still open whether \conn~is coNP-complete for every remaining
Horn set of relations, i.e. every set of Horn relations that contains
at least one relation that is not componentwise IHSB$-$.  Following
the same line of reasoning as in the proof of our Faithful
Expressibility Theorem we are able to show that one of the paths of
length 4 defined in Section \ref{sec:FET}, namely $M(\bar{x}_1,
\bar{x}_2, x_3)$, can be expressed faithfully from every such set of
relations. Thus the trichotomy would be established if one shows that 
\conn$(\{M(\bar{x}_1, \bar{x}_2, x_3) \})$ is \coNP-hard.

\section{Discussion and Open Problems}
\label{sec:disc}

In Section 2, we conjectured a trichotomy for
\conn$({\cal S})$.
%phk
 %We have made progress towards this conjecture, all that
%remains to establish a trichotomy is to pinpoint the complexity of
%\conn$({\cal S})$ when $S$ is Horn or dual Horn. We can extend our dichotomy
%theorem for  $st$-connectivity
%results on \stconn\ (Theorems \ref{thm:STCONN}
%and \ref{thm:diam})
%to formulas without constants, but this problem for  \conn$({\cal S})$ is
%still open.
In view of the results established here,  what remains is to
pinpoint the complexity of \conn$({\cal S})$ when ${\cal S}$ is
Horn but not componentwise IHSB$-$,  and when ${\cal S}$ is dual
Horn but not componentwise IHSB$+$.

 We can extend our dichotomy
theorem for $st$-connectivity
%results on \stconn\ (Theorems \ref{thm:STCONN}
%and \ref{thm:diam})
to \cnf$({\cal S})$-formulas without constants; the complexity of
connectivity for \cnf$({\cal S})$-formulas without constants is
open.
 %but this problem for  \conn$({\cal S})$ is
%still open.
We conjecture that when ${\cal S}$ is not tight, one can improve
the diameter bound from $2^{\Omega(\sqrt{n})}$ to $2^{\Omega(n)}$.
Finally,  we believe that our techniques can shed light on other
connectivity-related problems, such as approximating the diameter
and counting the number of components. \eat{For counting the number of
components, using results of Creignou and Hermann \cite{CH96}, we
can show that the problem is in \PTIME~ for affine, monotone and
dual monotone relations, and \#\PTIME-complete otherwise.}

\bibliographystyle{siam}
\bibliography{conn-csp}

\end{document}